\documentclass[pre,twocolumn,showpacs]{revtex4}
\usepackage{graphicx}
\begin{document}

\title{Matter-wave solitons in radially periodic potentials}
\author{Bakhtiyor B. Baizakov$^1$, Boris A. Malomed$^2$, and Mario Salerno$^3$\\
$^1$ Physical-Technical Institute of the Uzbek Academy of
Sciences, 2-b, G. Mavlyanov str., 700084, Tashkent, Uzbekistan \\
$^2$ Department of Interdisciplinary Studies, Faculty of
Engineering, Tel Aviv University,  Tel Aviv 69978, Israel  \\
$^3$ Dipartimento di Fisica "E.R. Caianiello", Consorzio Nazionale
Interuniversitario per le Scienze Fisiche della Materia (CNISM),
Universit\'{a} di Salerno, I-84081, Baronissi (SA), Italy}
%%\date{\today }

\begin{abstract}
We investigate two-dimensional (2D) states of Bose-Einstein
condensates (BEC) with self-attraction or self-repulsion, trapped
in an axially symmetric optical-lattice potential periodic along
the radius. Unlike previously studied 2D models with Bessel
lattices, no localized states exist in the linear limit of the
present model, hence all localized states are truly nonlinear
ones. We consider the states trapped in the central potential
well, and in remote circular troughs. In both cases, a new
species, in the form of \textit{radial gap solitons}, are found in
the repulsive model (the gap soliton trapped in a circular trough
may additionally support stable dark-soliton pairs). In remote
troughs, stable localized states may assume a ring-like shape, or
shrink into strongly localized solitons. The existence of stable
annular states, both azimuthally uniform and weakly modulated
ones, is corroborated by simulations of the corresponding
Gross-Pitaevskii equation. Dynamics of strongly localized solitons
circulating in the troughs is also studied. While the solitons
with sufficiently small velocities are stable, fast solitons
gradually decay, due to the leakage of matter into the adjacent
trough under the action of the centrifugal force. Collisions
between solitons are investigated too. Head-on collisions of
in-phase solitons lead to the collapse; $\pi $-out of phase
solitons bounce many times, but eventually merge into a single
soliton without collapsing.
\end{abstract}


\maketitle

\section{Introduction}

Matter-wave solitons, that can be created in Bose-Einstein
condensates (BECs), are a research subject of great interest, both
as nonlinear collective excitations in macroscopic quantum matter,
and due to the potential they offer in applications to
high-precision interferometry, atomic-wave soliton lasers, quantum
information processing, and other emerging technologies. Dark and
bright matter-wave solitons were experimentally created in nearly
one-dimensional (1D) trap configurations \cite{denschlag,
khaykovich, strecker}, filled with $^{87}$Rb and $^{7}$Li
condensate, respectively. In these settings, strong radial
confinement freezes transverse dynamics of the condensate, keeping
atoms (with mass $m$) in the ground state of the corresponding 2D
harmonic-oscillator potential, $m\omega _{\perp }^{2}r^{2}/2$,
while a weak axial parabolic trap, $m\omega _{\Vert }^{2}x^{2}/2$,
allows quasi-free motion of the soliton in the axial direction,
$x$. New recent experiments with $^{85}$Rb \cite{cornish} and
related numerical simulations \cite{parker} have showed the
existence of stable solitary waves in a weakly elongated ($\omega
_{\bot }/\omega _{x}\simeq 2.5$) trap. These essentially
three-dimensional (3D) bright solitons exhibit rich and complex
behavior, not present in their nearly-1D counterparts. Notable
among the new features, are inelastic collisions between solitons,
which are strongly sensitive to phase relations between them and
their relative speed.

Adding a periodic optical-lattice (OL) potential in the axial direction of
the quasi-1D trap makes it possible to create bright matter-wave solitons of
the gap type in repulsive condensates \cite{eiermann}, which is a particular
manifestation of very rich physics of BEC trapped in OLs, as reviewed in
Refs. \cite{morsch}.

Recently, 2D and 3D localized states in the Gross-Pitaevskii equation (GPE,
which is the basic mean-field BEC model), that may be sustained and
stabilized by cellular potentials, i.e., multidimensional OLs, have
attracted a great deal of interest, see review \cite{malomed1}. A great
challenge to the experiment is creation of multidimensional matter-wave
solitons, as well as the making of spatiotemporal solitons in nonlinear
optics \cite{malomed1}. In particular, multi-dimensional solitons trapped in
a low-dimensional OL (i.e., 1D lattice in the 2D space \cite{bms2004}, and
2D lattice in the 3D space \cite{bms2004,Barcelona}), have been predicted
\cite{bms2004}. As these solitons keep their mobility in the free direction,
the latter settings can be used to test head-on and tangential collisions
between solitons. Note that the quasi-1D lattice potential cannot stabilize
3D solitons \cite{bms2004} (this becomes possible if the quasi-1D lattice is
combined with the periodic time modulation of the nonlinearity provided by
the Feshbach-resonance-management technique \cite{Warszawa}; a general
account of the technique of periodic management for solitons can be found in
book \cite{book}).

The low-dimensional OL can also be implemented as a radial
(axially symmetric) lattice in both 2D
\cite{kartashov}-\cite{experiment} and 3D \cite{Bessel3D}
settings. In both cases, stable solitons in the self-attractive
medium were predicted, in the form of a spot trapped either at the
central potential well, or (in the 2D case) in a radial potential
trough (additionally, the so-called azimuthons were predicted as
azimuthally periodic deformations of vortices in the uniform 2D
medium \cite{azimuthon}; however, they are unstable in the case of
the cubic nonlinearity). In the latter case, the spot-shaped
soliton may run at a constant angular velocity in the trough (thus
being a \textit{rotary soliton} \cite{kartashov}). The circular
motion may be more convenient for the experimental study of
mobility and collisions of matter-wave solitons (and soliton
trains) than the already realized soliton-supporting settings in
the cigar-shaped traps \cite{khaykovich, strecker, parker}, as in
the circular geometry the motion is not affected by the weak
longitudinal trap. This setting is also convenient for studies of
patterns in self-repulsive BECs. In particular, dark solitons
were, thus far, created in cigar-shaped traps \cite{denschlag},
where the background density is essentially non-uniform, as it
vanishes at edges of the trap. In a circular trough, pairs of dark
solitons can be created without this complication. In fact, both
axisymmetric (azimuthally uniform) vortices \cite{repulsion} and
dipole and quadrupole states, that may be regarded as stable
complexes of two or four dark solitons \cite{Ricardo}, were
predicted in the two-dimensional GPE with the repulsive
nonlinearity and Bessel-lattice radial potential. The stable
dipole and quadrupole patterns may also rotate at a constant
angular velocity \cite{Ricardo}.

Radial potentials in the form of $-\varepsilon J_{0}\left(
kr\right) $ or $-\varepsilon J_{1}^{2}\left( kr\right) $, adopted
in Refs. \cite{kartashov} and \cite{repulsion,Ricardo},
respectively ($\varepsilon $ and $k$ are positive constants, and
$J_{0}$ and $J_{1}$ are the Bessel functions), can support
patterns of most types considered in those works in the linear
limit of the model. Indeed, axisymmetric states trapped in the
central potential well, as well as azimuthally uniform vortices
and dipolar and quadrupolar modes trapped in circular potential
troughs, may be found as 2D eigenstates of the quantum-mechanical
linear Schr\"{o}dinger equation in such radial potentials.
Accordingly, nonlinear states with the same symmetries found in
Refs. \cite{kartashov,repulsion,Ricardo} may be considered as
continuations of their linear counterparts. An exception is the
spot-shaped soliton in a circular trough (the one that can perform
rotary motion along the trough, as mentioned above). On the
contrary to that, in this paper we report results for the
\emph{radially periodic} OL potential, $-\varepsilon \cos (2kr)$,
which, obviously, cannot support any radially localized state in
the linear limit. Therefore, all localized patterns reported in
this paper are essentially nonlinear objects, and they may be
different from their counterparts found in the Bessel lattices, as
concerns their existence and stability conditions. The solitons
that we report in this work may be categorized as solitons of the
ordinary type in the model with self-attraction, and
\textit{radial gap solitons} (a new species of solitary waves), in
the case of self-repulsion. Note that solitons of the gap type
cannot exist in the above-mentioned Bessel-lattice potentials.

As for the attractive model, a somewhat similar problem was
considered in Refs. \cite{LC1} and \cite{LC2}, where it was
demonstrated that the self-attractive nonlinearity in the 1D and
2D geometry (the latter setting is axisymmetric) can turn a
quasi-bound state in a potential well dug on top of a potential
hill into a true bound state (including 2D vortices \cite{LC2}).
In the 3D setting, the situation is more complicated, but a
similar effect is also possible \cite{LC1}.

It is relevant to mention that, besides the attractive or repulsive cubic
nonlinearity, the radial-lattice potential can also be combined with the
self-focusing saturable nonlinearity, characteristic of photorefractive
media. It was predicted \cite{PLA} and very recently experimentally
demonstrated \cite{experiment} that stable solitons are possible in this
case too (in fact, Ref. \cite{experiment} presents the first ever
experimental realization of solitons in any radial lattice). Another very
recent experimental result is the observation of light localization, in the
form of spot-like and ring-like patterns, in a photorefractive medium
equipped with a third-order Bessel lattice subjected to azimuthal
modulation, which corresponds to a combination of saturable nonlinearity and
an effective potential $\sim J_{3}\left( r/r_{0}\right) \cos ^{3}\left(
3\theta \right) $, where $\theta $ is the angular coordinate \cite{n=3}.

A different experimental setup that also leads to a ring-shaped trap for the
BEC has been recently reported, based on a specially designed configuration
of the external magnetic field that creates a guide for matter waves in the
form of a torus \cite{torus-experiment}. In parallel, nearly-1D solitons in
toroidal traps loaded with self-attractive BEC were studied theoretically
\cite{torus-experiment} (in the experiment of Ref. \cite{torus-experiment},
the condensate was self-repulsive).

The objective of this paper is to investigate various types of
matter-wave solitons supported by the periodic radial lattice
potential. It should be mentioned that, besides the BEC context,
the model and results obtained in it may also be relevant to
photonic-crystal fibers with a concentric (rather than usual
hexagonal) structure, similar to that in fibers with the
multilayer cladding \cite{Fink}. In that case, the soliton will be
a self-trapped beam propagating in the fiber.

The paper is organized as follows. In Section II we formulate the
model based on the respective two-dimensional GPE. Section III is
dealing with solitons trapped in the center (including the radial
gap solitons, in the case of the model with repulsion), for which
results are obtained by means of the variational approximation
(VA; both static and dynamical variants of the VA are presented),
and in direct simulations. Section IV treats solitons trapped in
annular channels (potential troughs). By means of the VA, we
derive an effective 1D equation for the attractive model, which
predicts a possibility of the existence of stable axisymmetric
ring-shaped states, and the onset of their instability against
azimuthal modulations at a finite density. Exact modulated
(cnoidal-wave) solutions, which appear above the instability
threshold of the uniform state, are found too, as well as exact
solutions for azimuthal solitons. Direct numerical simulations
corroborate the existence of stable uniform and modulated
ring-shaped patterns, and reveal the radial gap solitons whose
central love is trapped in the annular trough (in the repulsive
version of the model). Also reported are stable ring-shaped gap
solitons carrying a pair of dark solitons on their crests. Section
V addresses dynamics of moving two-dimensional solitons, trapped
in the circular troughs of the attractive model. It is shown that
the slow solitons are stable, while fast ones are destroyed due to
leakage of matter into the outer trough, under the action of the
centrifugal force. Head-on collisions between in-phase solitons
lead to collapse (intrinsic blow-up), while $\pi $-out-of-phase
solitons collide many times, and eventually merge into a single
soliton (without collapsing); the same happens with in-phase
solitons colliding tangentially, as they are trapped in adjacent
troughs. Finally, Section VI concludes the paper.

\section{The model}

\begin{figure}[htb]
\centerline{\includegraphics[width=8cm,height=4cm,clip]{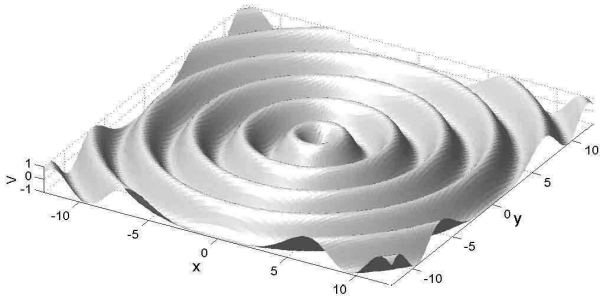}}
\caption{An example of the radially periodic axisymmetric
potential, $V(r)=-\cos (2r)$.} \label{fig1}
\end{figure}

The starting point is the standard GPE in a normalized form
\cite{morsch,book},
\begin{equation}
i\frac{\partial u}{\partial t}+\left( \frac{\partial
^{2}}{\partial r^{2}}+\frac{1}{r}\frac{\partial }{\partial
r}+\frac{1}{r^{2}}\frac{\partial ^{2}}{\partial \theta
^{2}}\right) u-V(r)u+\chi |u|^{2}u=0,  \label{gpe2D}
\end{equation}
written for the single-atom wave function $u$ in polar coordinates $\left(
r,\theta \right) $. Here, $\chi =+1$ and $-1$ correspond to the
self-focusing and defocusing nonlinearity, respectively (alias
negative and positive scattering length of interatomic
collisions), and, as said above, $V(r)=\varepsilon \cos (2kr)$ is
the radially periodic potential with strength $\varepsilon $ and
wavenumber $2k$ (by dint of obvious rescaling, we set $k\equiv
1$). In terms of OLs, this potential can be created by a
cylindrical beam whose amplitude is modulated as $\cos (kr)$ (see
further discussion below). Note that the potential with
$\varepsilon >0$ gives rise to a local minimum at $r=0$, see an
example in Fig. \ref{fig1}.

Together with Eq. (\ref{gpe2D}), we will use its variational
representation. Obviously, the equation can be derived from the
Lagrangian
\begin{eqnarray} L &=&\int_{0}^{\infty
}rdr\int_{0}^{2\pi }d\theta \left[ \frac{i}{2}\left(
\frac{\partial u}{\partial t}u^{\ast }-\frac{\partial u^{\ast }}
{\partial t}u\right) -\right.  \nonumber \\
&&\left. \left( \left\vert \frac{\partial u}{\partial
r}\right\vert ^{2}+\frac{1}{r^{2}}\left\vert \frac{\partial
u}{\partial \theta }\right\vert ^{2}\right)
+V(r)|u|^{2}+\frac{1}{2}\chi |u|^{4}\right] ,  \label{Lagr}
\end{eqnarray}
where $\ast $ stands for the complex conjugate.
Stationary axially symmetric states with chemical potential $\mu $
are looked for as solutions to Eq. (\ref{gpe2D}) in the form of
$u=e^{-i\mu t}U(r)$, with a real function $U$ obeying the
equation\begin{equation}
\mu U+\left(
\frac{d^{2}}{dr^{2}}+\frac{1}{r}\frac{d}{dr}\right) U-V(r)U+\chi
U^{3}=0.  \label{U}
\end{equation}
For given potential $V(r)$, solutions to Eq. (\ref{U}) form families
parameterized by $\mu $ or, alternatively, by the
norm,\begin{equation} N=2\pi \int_{0}^{\infty }\left( U(r)\right)
^{2}rdr\equiv N=2\pi \int_{0}^{\infty }|u|^{2}rdr.  \label{Norm}
\end{equation}

Solutions to stationary equation (\ref{U}) with $\chi =+1$ (self-attraction)
were obtained by means of a recently developed spectral-renormalization
method for finding self-localized states of nonlinear-Schr\"{o}dinger (NLS)
type equations \cite{am}. Independently, the same stationary solutions were
also generated by dint of the well-known method which uses the integration
of GPE (\ref{gpe2D}) in imaginary time \cite{chiofalo}. Stability of the
solitary waves was then tested by adding small perturbations to them and
integrating the GPE in real time. A stable perturbed soliton would shed off
some radiation, which was absorbed at boundaries of the integration domain,
and relax into a slightly different stationary form, corresponding to the
smaller norm.

In the case of repulsion ($\chi =-1$), self-localized states were obtained
by solving Eq. (\ref{gpe2D}) in real time, with an absorber placed at the
boundary of the integration domain. In that case, a localized waveform
(e.g., Gaussian) with a suitable norm was chosen as the initial condition.
In the course of the evolution, excess norm is radiated away with linear
waves, which are absorbed at the boundary, and a self-localized state
emerges (see an example in Fig. \ref{fig4} below). It is obvious that
solitons found this way are stable.

The above-mentioned radial lattices of the Bessel type can be
naturally generated by a nondiffracting linear optical beam with
the cylindrical symmetry (Bessel beams themselves are usually
generated by means of axicoms) \cite{Durnin}. On the other hand,
the radial potential of the form $\cos (2kr)$ can be induced, as
said above, by a cylindrical beam whose amplitude is modulated as
$\cos (kr)$. The modulation can be provided by passing the
cylindrical laser beam through a properly shaped plate
\cite{PLA,experiment}. Unlike the Bessel beam, the one with the
$\cos (kr)$ transverse modulation will suffer conical diffraction,
but it is not a problem for BEC\ experiments, as a tight optical
trap created in the transverse direction may readily contain a 2D
pancake-shaped configuration of the condensate, of the thickness
$\simeq 2$ $\mu $m \cite{layer}. The diffraction of the paraxial
beam, with waist diameter $D\sim 100$ $\mu $m, relevant to the
experiment, on such a short propagation distance is completely
negligible, as the respective diffraction length (alias Rayleigh
range) is estimated to be $z_{\mathrm{diffr}}\sim D^{2}/\lambda
\sim 10$ mm, where $\lambda \sim 1$ $\mu $m is the beam's carrier
wavelength.

\section{Solitons trapped at the center}

\subsection{Numerical solutions}

\begin{figure}[tbh]
\centerline{\includegraphics[width=8cm,height=6cm,clip]{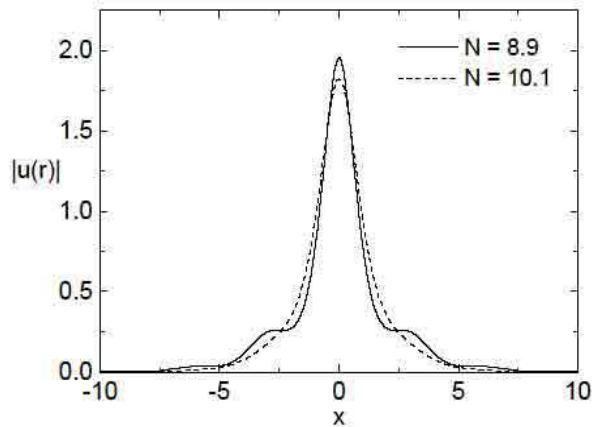}}
\caption{Axial cross sections of two stable solitons with equal
chemical potentials, $\protect\mu =-0.5$, trapped (in the model
with self-attraction) at the center of the radial structure,
$V(r)=\protect\varepsilon \cos (2r)$, with $\protect\varepsilon
=1.5$ (in this case, the norm is $N=8.9$; solid line), and
$\protect\varepsilon =0.5$ ($N=10.1$, dashed line).} \label{fig2}
\end{figure}

\begin{figure}[tbh]
\centerline{\includegraphics[width=8cm,height=4cm,clip]{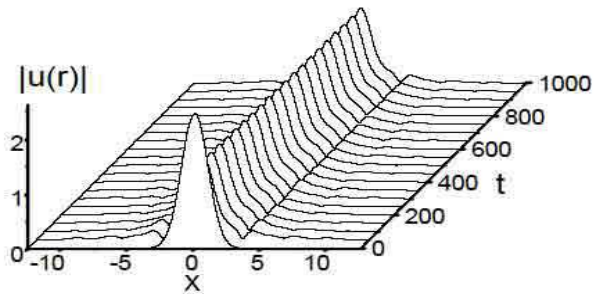}}
\caption{Formation of a self-trapped state (\textit{radial gap
soliton}) in repulsive BEC in the radial lattice with strength
$\protect\varepsilon =3$ (in the presence of absorbers at the
boundary of the integration domain) from an initial Gaussian with
norm $N=6\protect\pi $. The norm of the eventually established
state is $N=5.63$, i.e., $\approx 30\%$ of the initial value.}
\label{fig3}
\end{figure}

\begin{figure}[tbh]
\centerline{
\includegraphics[width=4cm,height=3cm,clip]{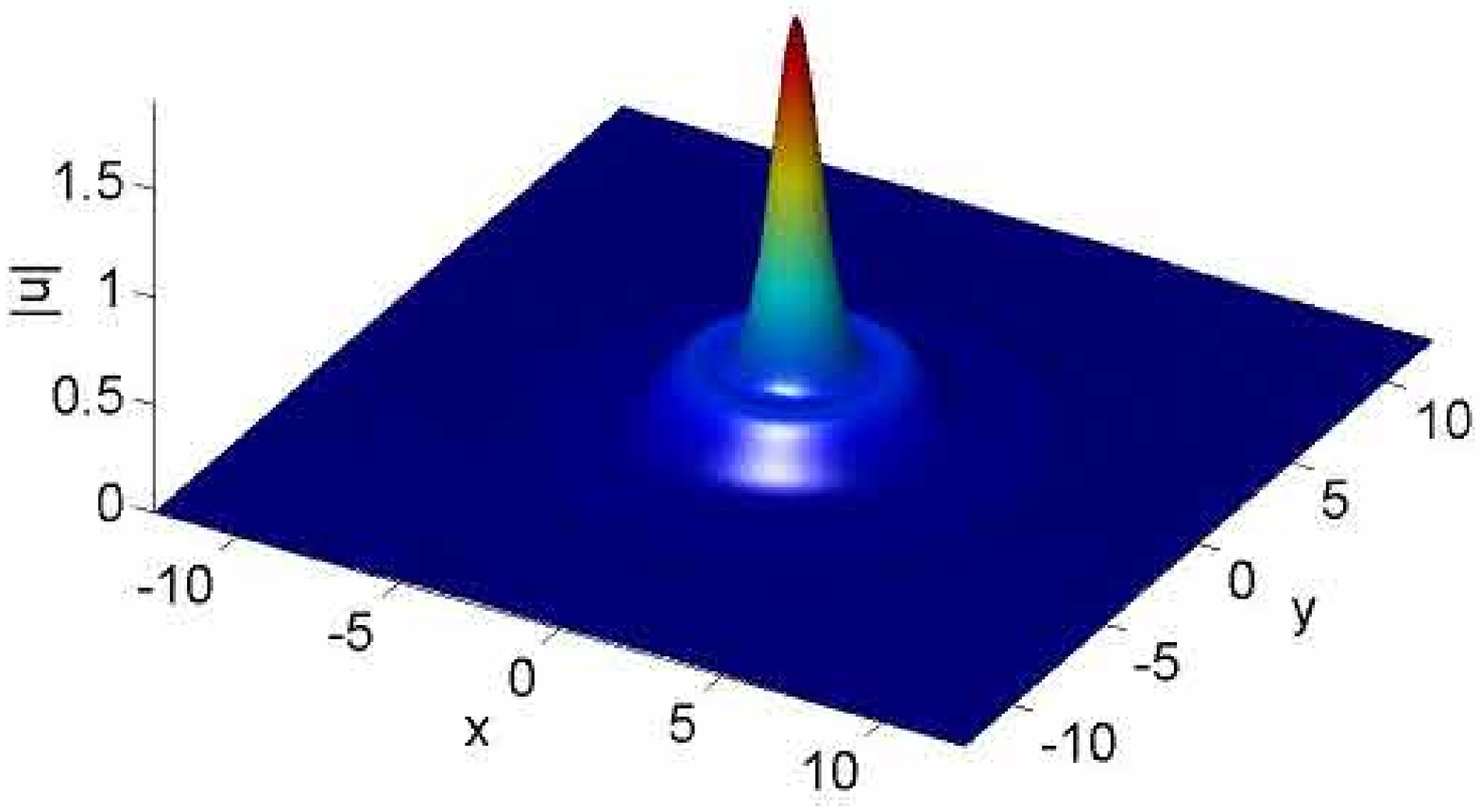} \quad
\includegraphics[width=4cm,height=3cm,clip]{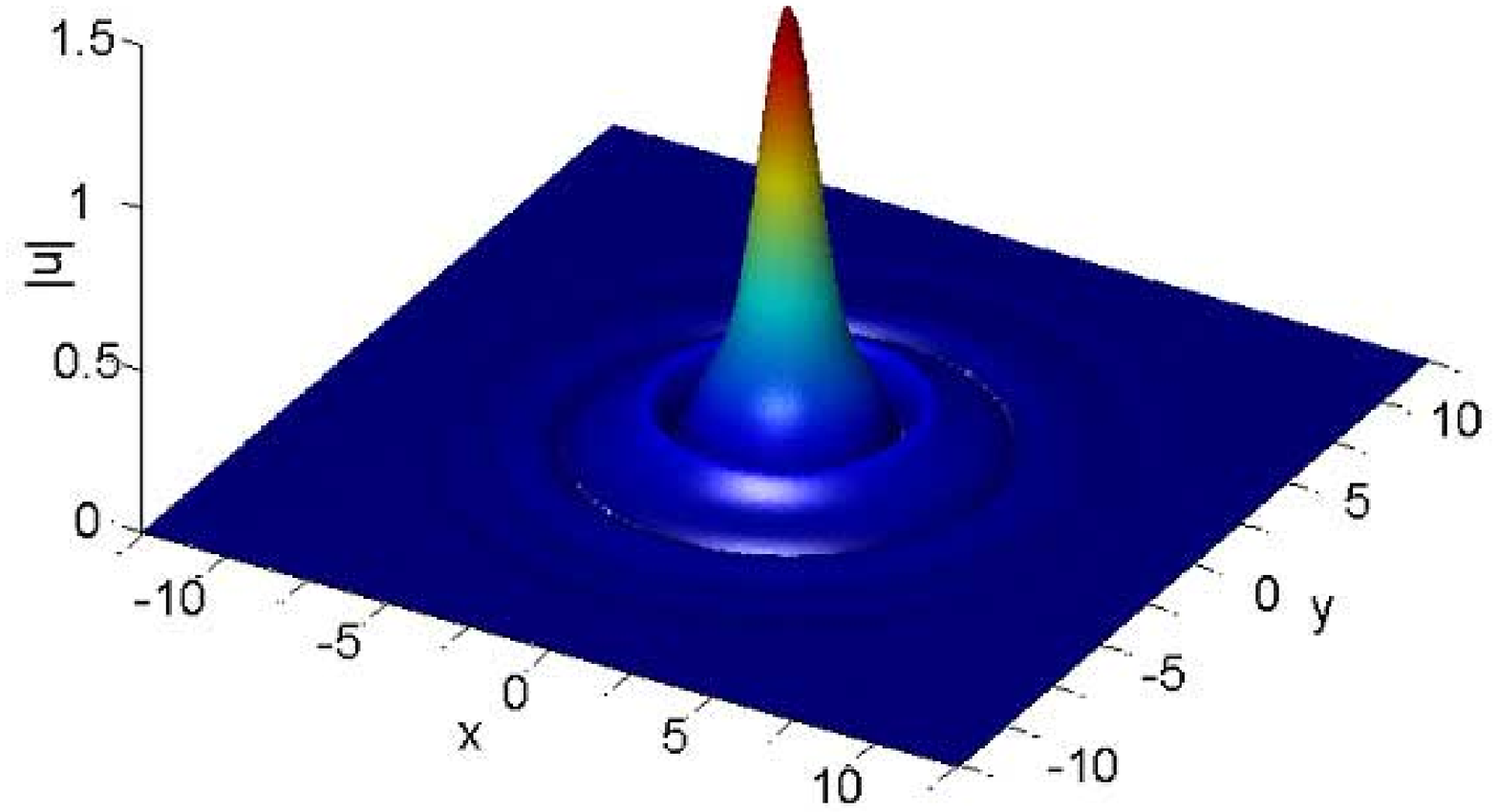}
} \caption{Left panel: The same localized state in the attractive
BEC whose axial cross section is shown by the solid curve in Fig.
\protect\ref{fig2} (with $N=8.9$, $\protect\mu =-0.5$, and
$\protect\varepsilon =1.5$). Right panel: A localized state in the
repulsive BEC, with $N=5.63$, $\protect\mu =1.88$, amplitude
$A=1.358$, and $\protect\varepsilon =3$, which was generated as
shown in Fig. \protect\ref{fig3}.} \label{fig4}
\end{figure}

It is known that a radial potential structure can easily trap a
soliton in the central potential well \cite{kartashov}. We start
the consideration of the present model with solutions of this
type. Typical examples of the solitonic shape in the model with
self-attraction ($\chi =+1$) are displayed in Fig. \ref{fig2}.
These solutions were obtained by means of both the above-mentioned
spectral-renormalization method \cite{am} and integration in
imaginary time \cite{chiofalo}.
In the repulsive model ($\chi =-1$), numerical solutions were
obtained, as said above, by solving Eq. (\ref{gpe2D}) in real
time, in the presence of the boundary absorber. An example of the
relaxation of an initial Gaussian pulse into a soliton is given in
Fig. \ref{fig3}. Full views of the typical stable solitons trapped
at the center of the radial structure in the model with
self-attraction and self-repulsion are displayed in Fig.
\ref{fig4}.

As concerns the solitons in the repulsive model ($\chi =-1$), a
typical example of which is displayed in the right panel of Fig.
\ref{fig4}, they may be clearly classified as solitons of the gap
type, since they exist solely due to the interplay of the lattice
potential and self-defocusing nonlinearity. The ordinary setting
that gives rise to gap solitons is quite similar, with the OL in
the 1D, 2D or 3D Cartesian coordinates \cite{GS}. Slowly decaying
fringes, attached to the central core of the soliton and evident
in the right panel of Fig. \ref{fig4} (and absent in the ordinary
soliton displayed in the left panel), is a characteristic feature
of gap solitons. Thus, this self-trapped state may be naturally
called a \textit{radial gap soliton}.

\subsection{Variational approximation}

To describe a stationary soliton trapped in the central potential well in an
analytical approximation, we adopt a Gaussian ansatz,
\begin{equation}
U(r)=A\exp \left( -\frac{r^{2}}{2a^{2}}\right) .  \label{ansatz-zentrum}
\end{equation}
Comparison with Fig. \ref{fig2} shows that, in the model with
attraction ($\chi =+1$), this ansatz is quite adequate for
solitons with a larger norm, trapped in a relatively weak lattice,
but it may be inaccurate as an approximation for solitons with
smaller $N$ trapped in a stronger lattice.

The application of the variational approximation (VA) to
stationary equation (\ref{U}), with the help of the accordingly
modified Lagrangian (\ref{Lagr}), is straightforward (cf. Refs.
\cite{bms2003,bms2004,Estoril}, where the same Gaussian ansatz was
used to predict 2D solitons trapped in quasi-1D and square OLs).
Thus we arrive at a set of VA-generated equations that relate the
soliton's size $a$ to its norm and chemical potential:
\begin{equation}
\frac{\chi N}{4\pi }=1-\frac{1}{2}\varepsilon a^{3}\frac{df}{da},
\quad \mu =\varepsilon \left[ a\frac{df}{da}+f(a)\right]
-\frac{1}{a^{2}},  \label{va1}
\end{equation}
where $f(a)=1-\sqrt{\pi }ae^{-a^{2}}\mathrm{erfi}(a)$,
$\mathrm{erfi}(a)\equiv \mathrm{erf}(ia)/i$, and $\mathrm{erf\ }$is the standard error
function.

\begin{figure}[tbh]
\centerline{\includegraphics[width=8cm,height=6cm,clip]{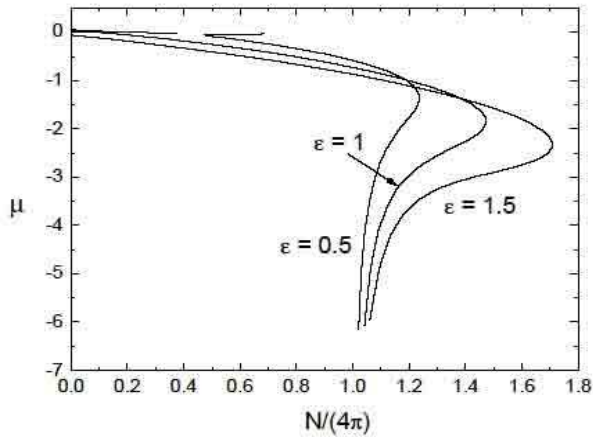}}
\caption{Dependences $\protect\mu =\protect\mu (N)$ following from
variational equations (\protect\ref{va1}), for the self-attraction
case, $\protect\chi =+1$. According to the VK criterion,
$d\protect\mu /dN<0$, stable solutions belong to upper branches of
the curves.} \label{fig5}
\end{figure}

The analysis of dependence $\mu=\mu(N)$ following from Eq.
(\ref{va1}) with $\chi =+1$ (self-focusing) predicts the existence
of a family of solitons trapped at the center of the radial
lattice, which are stable according to the Vakhitov-Kolokolov (VK)
stability criterion \cite{VK,berge}, $d\mu /dN<0$, as shown in
Fig. \ref{fig5}. Although the VK criterion is only necessary for
the stability, as it ignores a possibility of oscillatory
instability with complex eigenvalues, direct simulations of Eq.
(\ref{gpe2D}) confirm the stability of these solutions (see
below).

\begin{figure}[tbh]
\centerline{
\includegraphics[width=4cm,height=4cm,clip]{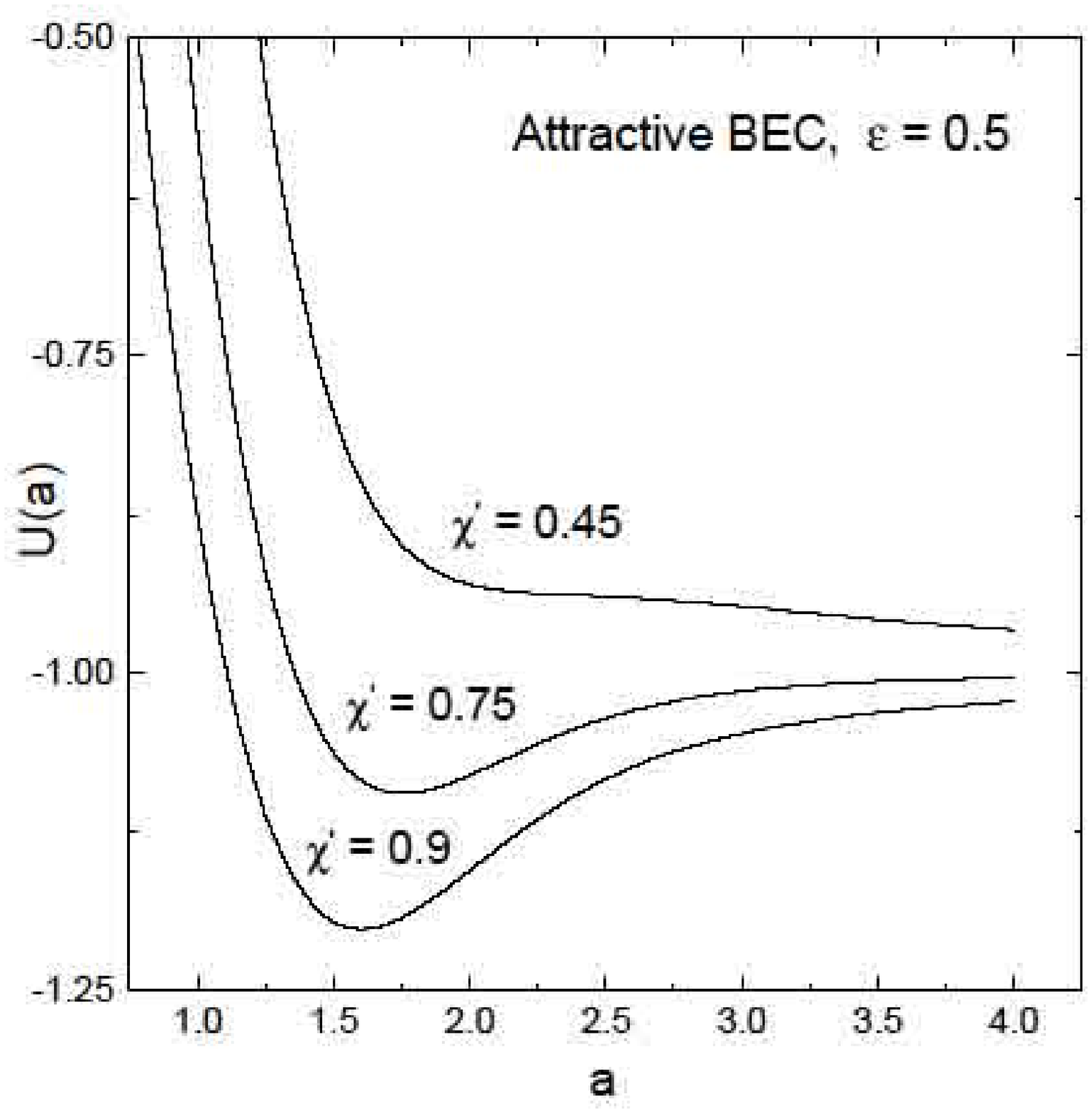} \quad
\includegraphics[width=4cm,height=4cm,clip]{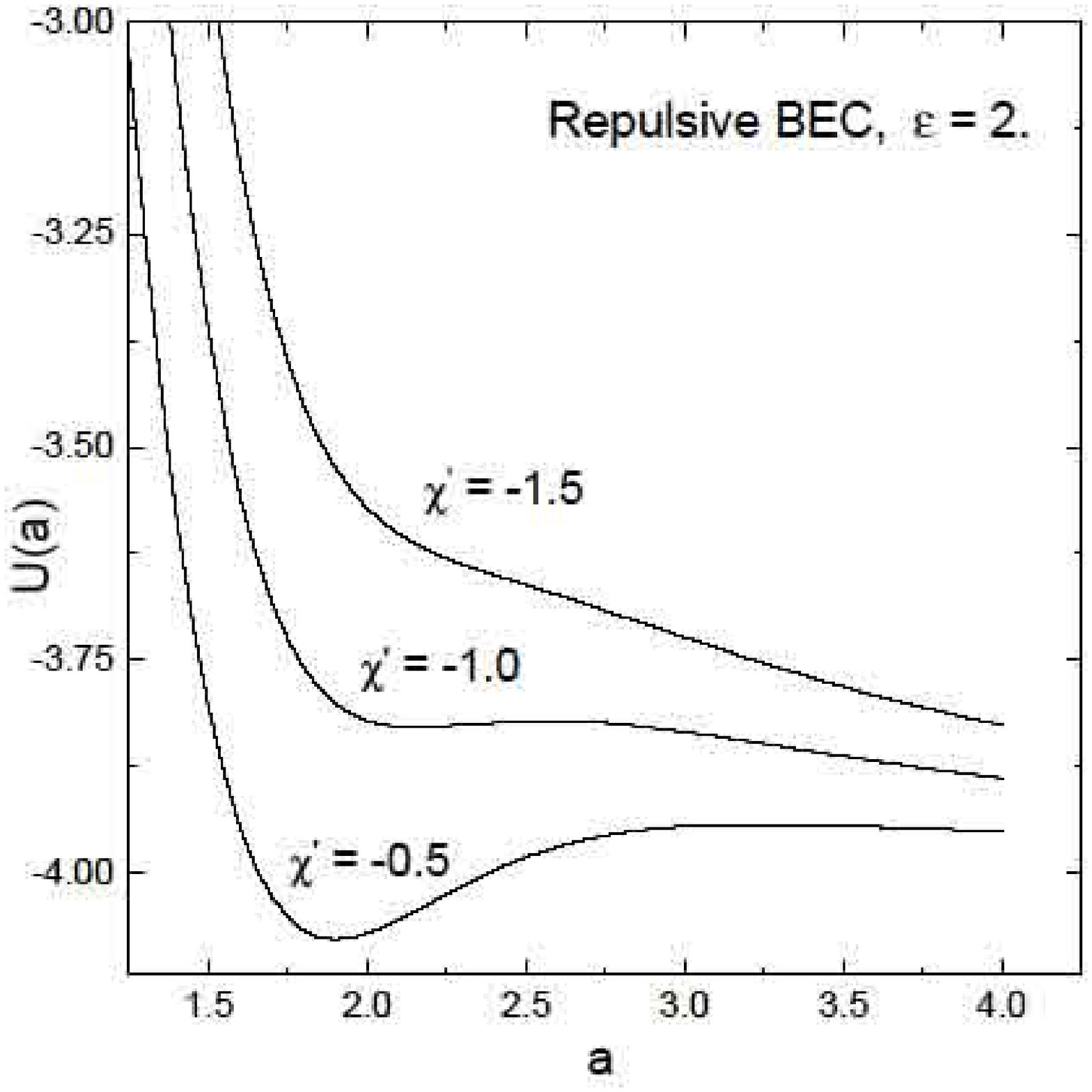}
} \caption{The variational potential, as given by Eq.
(\protect\ref{pot}) for different values of $\protect\chi
^{^{\prime }}=\protect\chi N/(4\protect\pi )$, at fixed strength
$\protect\varepsilon $ of the radial optical lattice.}
\label{fig6}
\end{figure}

Additional evidence for the existence of self-trapped localized
states of BEC in the present model can be provided by a more
general, time-dependent, version of the VA. To this end, we define
a generalized ansatz [cf. Eq. (\ref{ansatz-zentrum})]
\begin{equation}
u(r,t)=A(t)\exp \left[ -\frac{r^{2}}{2\left( a(t)\right)
^{2}}+\frac{i}{2}b(t)r^{2}+i\phi (t)\right] ,  \label{ansatz}
\end{equation}
where $b\ $is a real radial chirp. Applying the standard VA procedure
\cite{malomed2}, one arrives at the following evolution equation for the width of
the localized state,
\begin{equation}
\frac{d^{2}a}{dt^{2}}=\frac{4(1-\chi ^{\prime })}{a^{3}}+2\varepsilon \left[
2a+\sqrt{\pi }(1-2a^{2})e^{-a^{2}}\mathrm{erfi}(a)\right] ,  \label{att}
\end{equation}
where an effective nonlinearity strength is
\begin{equation}
\chi ^{\prime }\equiv \chi N/(4\pi ),  \label{chi}
\end{equation}
and the amplitude is given by
\begin{equation}
A^{2}(t)=N/\left( \pi a^{2}(t)\right) .  \label{Ampl}
\end{equation}
Equation (\ref{att}) describes motion of a unit-mass particle with
the coordinate $a(t)$ in an effective potential
\begin{equation}
U(a)=\frac{2(1-\chi ^{\prime })}{a^{2}}-2\sqrt{\pi }\varepsilon
ae^{-a^{2}}\mathrm{erfi}(a).  \label{pot}
\end{equation}
In Fig. \ref{fig6} the effective potential (\ref{pot}) is depicted
for different values of $\chi ^{\prime }$, for the attractive and
repulsive versions of the model.

Solitons in the attractive BEC ($\chi =+1$) exist when potential
(\ref{pot}) possesses a local minimum. In particular, for
$\varepsilon =0.5$ (see Fig. \ref{fig6}), the potential minimum
exists provided that $\chi ^{\prime }>0.45 $. For the norm smaller
than that corresponding to $\chi ^{\prime }=0.45$ [see Eq.
(\ref{chi})], the self-attraction is too weak to form a soliton,
and the matter-wave pulse spreads out through barriers of the
radial potential. On the other hand, in the case of attraction,
$\chi ^{\prime }$ is also limited from above, by the collapse
occurring at a critical value, $\chi _{\mathrm{cr}}^{\prime
}=N_{\mathrm{cr}}/(4\pi )$. In the absence of the lattice
($\varepsilon =0$), a numerically exact critical norm is
$N_{\mathrm{cr}}^{(0)}\simeq 11.7$ (it is equal to the norm of the
\textit{Townes soliton}) \cite{berge}. For any $\varepsilon $, the
VA predicts $\chi _{\mathrm{cr}}^{\prime }=1$, as the value of
coefficient $\chi ^{\prime }$ at which the first term in effective
potential (\ref{pot}) vanishes. According to Eq. (\ref{chi}), this
gives $N_{\mathrm{cr}}^{(\mathrm{var})}=4\pi $ \cite{Sweden}.

Thus, the VA predicts a finite existence region for stable
self-trapped localized states in the attractive BEC; for example,
it is $0.45<\chi ^{\prime }<1$ for $\varepsilon =0.5$. While the
upper edge of this region is fixed at
$N=N_{\mathrm{cr}}^{(\mathrm{var})}=4\pi $, numerical analysis of
expression (\ref{pot}) demonstrates that the lower critical norm
necessary for the existence of the self-trapped state almost
linearly decreases with the increase of the OL strength
$\varepsilon $. Numerically exact solution of the $\theta
$-independent version of full GPE (\ref{gpe2D}) demonstrates that,
although the collapse sets in at a critical value of the norm
which is slightly higher than
$N_{\mathrm{cr}}^{(\mathrm{var})}=4\pi $, similar to the case of
the 2D soliton stabilized by the quasi-1D or square OL
\cite{bms2003,bms2004,Estoril}, the dependence of
$N_{\mathrm{cr}}$ on $\varepsilon $ is weak. The physical
mechanism behind the rise of the collapse threshold in a trapping
potential is that the confinement allows the soliton to increase
the internal ``quantum pressure", generated by the dispersion term
in GPE (\ref{gpe2D}), which counteracts the nonlinear
self-focusing of the wave packet.

\begin{figure}[tbh]
\centerline{\includegraphics[width=6cm,height=4cm,clip]{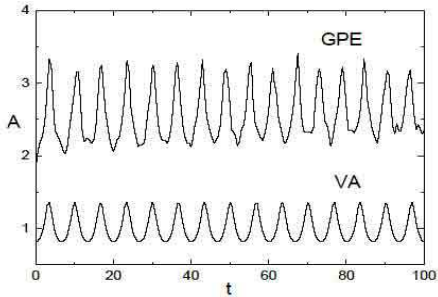}}
\caption{Comparison of a result of the variational approximation
(``VA") with direct simulations of Eq. (\protect\ref{gpe2D})
(``GPE") with $\protect\chi =+1$, in terms of the soliton's
amplitude as a function of time. Intrinsic vibrations of the
soliton were triggered by sudden increase of the lattice strength
from $\protect\varepsilon =0.5$ to $\protect\varepsilon =0.75$,
for the self-trapped state with $N=10.1$ in Fig.
\protect\ref{fig2}.} \label{fig7}
\end{figure}

Predictions of the VA for dynamical regimes can be compared to
direct simulations in terms of the time dependence of the
amplitude, $A(t)$, in a soliton which was excited by a sudden
perturbation [the variational prediction for $A(t)$ is taken as
per Eq. (\ref{Ampl})]. To display an example, we take the
stationary solution from Fig. \ref{fig2}, corresponding to
$\varepsilon =0.5$ and $N=10.1$, and suddenly increase the lattice
strength to $\varepsilon =0.75$. Comparison of the variational
results with simulations of Eq. (\ref{gpe2D}) is presented in Fig.
\ref{fig7}. As observed, agreement for absolute values of the
amplitude is, at best, qualitative, while the oscillation
frequencies coincide much better.

In the case of the repulsive BEC ($\chi =-1$), the behavior is
opposite, as seen in the right panel of Fig. \ref{fig6}. In this
case, a larger norm for given OL strength $\varepsilon $ causes
spreading of the localized state, because the self-defocusing
nonlinearity overcomes the confining effect of the trapping
potential. Simulations of Eq. (\ref{gpe2D}) with $\chi =-1$ show
that the wave packet sheds off excessive norm in the form of
linear waves, which are absorbed at domain boundaries, and relaxes
into a stable shape after the norm becomes adequate. For example,
at $\varepsilon =2$ stable localized states in the repulsive BEC
exist at $|\chi ^{\prime }|$ $_{\sim }^{<}~1$ [i.e., the norm is
limited to values $N$ $_{\sim }^{<}~4\pi $, see Eq. (\ref{chi})].

\section{Ring-shaped solitons}

\subsection{Analytical approach}

The most essential property of radially periodic potentials is that
localized states can self-trap not only in the central potential well, but
also in remote circular troughs \cite{kartashov}. If the trough's curvature
is not significant, these localized states resemble 2D solitons in the
quasi-1D OL reported in Ref. \cite{bms2004}. Here, we aim to derive an
asymptotic 1D equation for patterns trapped in a circular trough, and obtain
its relevant solutions. To this end, we adopt the following ansatz for the
solution,
\begin{equation}
\psi (r,\theta ,t)=\sqrt{2}A(y,t)\mathrm{sech}\left( x/\rho \right) ,
\label{ansatz1}
\end{equation}
where $y\equiv r_{0}\theta $ is the coordinate running along the
circumference of radius $r_{0}$, $A(y,t)$ is a slowly varying
complex amplitude, and the transverse variable is $x\equiv r-\pi
m$, with $r_{0}\equiv \pi m$ being a radial potential minimum in a
vicinity of which the ring-shaped pattern is trapped. Assuming $m$
to be large enough, we consider the annular pattern placed in a
remote circular trough, which may be treated as a
quasi-rectilinear potential trap, with negligible curvature.
Obviously, the latter condition amounts to $\rho \ll r_{0}$
[recall $\rho $ is the radial size in ansatz (\ref{ansatz1})].

Neglecting the curvature in Eq. (\ref{gpe2D}), we rewrite it as a 2D
equation with $x$ and $y$ treated as local Cartesian coordinates,
\begin{equation}
i\frac{\partial \psi }{\partial t}+\frac{\partial ^{2}\psi
}{\partial x^{2}}+\frac{\partial ^{2}\psi }{\partial y^{2}}+\chi
\left\vert \psi \right\vert ^{2}\psi -\varepsilon \cos \left(
2x\right) \psi =0.  \label{no-curv}
\end{equation}
The next step is to reduce Eq. (\ref{no-curv}) to an effectively 1D equation
along azimuthal direction $y$. To this end, we note that Eq. (\ref{no-curv})
corresponds to the Lagrangian [cf. Eq. (\ref{Lagr})]
\begin{eqnarray}
L=\int\int dxdy && \left[ \frac{i}{2}\left( \psi _{t}\psi ^{\ast
}-\psi \psi _{t}^{\ast }\right) -\left( \left\vert \psi
_{x}\right\vert ^{2}+\left\vert \psi _{y}\right\vert ^{2}\right)
\right.\nonumber \\
&& \left. +\frac{1}{2}\chi \left\vert \psi \right\vert
^{4}-\varepsilon \cos (2x)\left\vert \psi \right\vert ^{2}\right].
\end{eqnarray}
In this Lagrangian, we substitute ansatz (\ref{ansatz1}) and
integrate the resulting expression in transverse direction $x$.
After a straightforward calculation, this yields an effective
(averaged) Lagrangian,
\begin{eqnarray}
L_{\mathrm{eff}} &=& 2\int_{-\infty }^{+\infty }dy\left[ i\rho
\left( A_{t}A^{\ast }-AA_{t}^{\ast }\right) -2\rho \left\vert
A_{y}\right\vert ^{2} -2I\frac{\rho _{y}^{2}}{\rho }|A|^{2}
\right. \nonumber \\
&& \left.
-\frac{2\left\vert A\right\vert ^{2}}{3\rho
}+\frac{4}{3}\chi \rho \left\vert A\right\vert ^{4} -\frac{2\pi
\varepsilon \rho ^{2}\left\vert A\right\vert ^{2}}{\sinh \left(
\pi \rho \right) }\right] , \label{Leff}
\end{eqnarray}
where $I\equiv \int_{0}^{+\infty }\left(
\mathrm{sech}^{2}x-\mathrm{sech}^{4}x\right) x^{2}dx\approx 0.61$.

The application of the standard variational procedure to averaged
Lagrangian (\ref{Leff}) gives rise to a cumbersome system of
coupled equations for the complex amplitude $A(y,t)$ and real
width $\rho (y,t)$. The equations strongly simply if we assume
that $A$ and $\rho $ are slowly varying functions of $y$, which,
in particular, corresponds to long-wave perturbations that account
for the onset of the modulational instability (MI), see below. In
particular, in the attractive model ($\chi =+1$) one may first
adopt the lowest-order approximation, completely neglecting the
$y$-dependence and assumes time-independent $|A|$ and $\rho $,
which gives rise to a result previously known from the application
of the VA to the NLS equation in 1D with potential $\cos (2kx)$
\cite{Wang}$\allowbreak $:

\begin{equation}
|A|=\sqrt{\frac{1}{\rho ^{2}}+\frac{3}{2}\varepsilon \pi \rho \left[ \pi
\rho \frac{\cosh (\pi \rho )}{\sinh ^{2}(\pi \rho )}-\frac{1}{\sinh \left(
\pi \rho \right) }\right] }.  \label{A^2}
\end{equation}
This relation is cumbersome too, but straightforward analysis demonstrates
that it takes a very simple form, $|A|=1/\rho $, when $\varepsilon $ is both
small and large [in the latter case, scaling $|A|\sim \rho ^{-1}\sim
\varepsilon ^{-1/2}$ is assumed, to lend the last three term in the
Lagrangian density in Eq. (\ref{Leff}) the same order of magnitude].

To obtain a closed-form Lagrangian for the amplitude field
$A(y,t)$ in the attractive model, one should express $\rho $ in
Eq. (\ref{Leff}) in terms of $\left\vert A\right\vert $, using
relation (\ref{A^2}). This will lead, in the general case, to a
messy result; however, the above-mentioned simplest approximation,
$\rho =1/\left\vert A\right\vert $, which replaces Eq.
(\ref{A^2}), yields a tractable expression,
\begin{eqnarray}
L_{\mathrm{eff}} &=& 2\int_{-\infty }^{+\infty }dy \left\{ i\left(
A_{t}\sqrt{\frac{A^{\ast }}{A}}-A_{t}^{\ast
}\sqrt{\frac{A}{A^{\ast }}}\right) +\frac{4}{3}\left\vert
A\right\vert ^{3}
\right.  \nonumber \\
&&\left. -\frac{I}{2}\left[ \frac{\left( A_{y}^{\ast }\right)
^{2}}{A^{\ast }}\sqrt{\frac{A}{A^{\ast
}}}+\frac{A_{y}^{2}}{A}\sqrt{\frac{A^{\ast }}{A}}\right]
\right.  \nonumber \\
&&\left. -\left( 2+I\right) \frac{\left\vert A_{y}\right\vert
^{2}} {\left\vert A\right\vert} -2\varepsilon |A| \right\}
\label{Lfinal}
\end{eqnarray}
[the last term in the Lagrangian density should be dropped if
$\varepsilon $ is small; if $\varepsilon $ is large, the term is
obtained within the framework of the above scaling, that assumes
small $\rho $, hence $\sinh (\pi \rho )\approx \pi \rho $].
Finally, the action functional, $S=\int L_{\mathrm{eff}}dt$, as
defined by Lagrangian (\ref{Lfinal}), gives rise to the
Euler-Lagrange equation, $\delta S/\delta A^{\ast }=0$, in the
following form:
\begin{eqnarray}
&& i\tilde{A}_{t}+(2+I)\tilde{A}_{yy}-\left( 1+\frac{3I}{4}\right)
\frac{\tilde{A}_{y}^{2}}{\tilde{A}}+I\left[
\frac{\tilde{A}}{\tilde{A}^{\ast }}\tilde{A}_{yy}^{\ast }
\right.  \nonumber \\
&&\left. +\frac{\left\vert \tilde{A}_{y}\right\vert
^{2}}{2\tilde{A}^{\ast }} -\frac{3}{4}\frac{\left(
\tilde{A}_{y}^{\ast }\right) ^{2}\tilde{A}}{\left( \tilde{A}^{\ast
}\right) ^{2}}\right] +2\left\vert \tilde{A}\right\vert
^{2}\tilde{A}=0,  \label{new}
\end{eqnarray}
where $\tilde{A}(y,t)\equiv A(y,t)e^{-i\varepsilon t}$ [this transformation
eliminates a linear term in the equation generated by the last term in
Lagrangian (\ref{Lfinal})].

Formally setting $I=0$ in Eq. (\ref{new}) recovers an equation known as a
model for the propagation of surface waves on a plasma layer with a sharp
boundary \cite{Gradov}. A class of nonpolynomial generalized NLS equations
similar to Eq. (\ref{new}), and some of their fundamental solutions, such as
single-soliton ones, were introduced in Ref. \cite{Lennart} (see also Ref.
\cite{Nattermann}).

Stationary solutions to Eq. (\ref{new}) with chemical potential
$\mu $ are looked for in the ordinary form,
$\tilde{A}(y,t)=\allowbreak e^{-i\mu t}B(y)$, with $B(y)$
satisfying equation $(1+I)\left[ 2B^{\prime \prime }-(B^{\prime
})^{2}/B\right] +2B^{3}+\mu B=0$, where $B^{\prime }\equiv dB/dy
$.$\allowbreak $ This equation has its Hamiltonian,
\begin{equation}
H=(2/3)B^{3}+\mu B+(1+I)(B^{\prime })^{2}/B,  \label{H}
\end{equation}
Setting $H=0$, one obtains a family of soliton solutions, for any $\mu <0$:
\begin{equation}
B_{\mathrm{sol}}(y)=\sqrt{-\frac{3\mu }{2}}\mathrm{sech}\left(
\sqrt{-\frac{\mu }{1+I}}y\right) .  \label{soliton}
\end{equation}
For $H<0$ and $\mu <0$, a family of periodic cnoidal-wave solutions is
obtained in the form
\begin{equation}
B_{\mathrm{cn}}(y)=\frac{B_{0}B_{1}}{B_{1}+\left(
B_{0}-B_{1}\right) \mathrm{cn}^{2}\left( \sqrt{\frac{B_{0}\left(
B_{1}+\left\vert B_{2}\right\vert \right) }{6\left( 1+I\right)
}}y,q\right) },  \label{cn}
\end{equation}
where $\mathrm{cn}(z,q)$ is the Jacobi's elliptic cosine, with modulus
\begin{equation}
q=\sqrt{\frac{B_{0}-B_{1}}{B_{0}}\cdot \frac{\left\vert
B_{2}\right\vert }{B_{1}+\left\vert B_{2}\right\vert }}<1,
\label{modulus}
\end{equation}
and $B_{2}<0<B_{1}<B_{0}$ are three roots of equation $(2/3)B^{3}+\mu B-H=0$
[it is obtained by setting $B^{\prime }=0$ in Eq. (\ref{H})].
Solutions (\ref{cn}) exist, for given $\mu <0$, in a region of
$0\leq -H\leq \left( \sqrt{2}/3\right) |\mu |^{3/2}$; as said
above, the limit of $H=0$ corresponds to soliton (\ref{soliton}),
and the opposite limit, $H=-\left( \sqrt{2}/3\right) |\mu
|^{3/2}$, corresponds to a uniform CW (continuous-wave) solution,
with
\begin{equation}
B_{0}=B_{1}\equiv B(y)=\sqrt{|\mu |/2},~B_{2}=-\sqrt{2|\mu |}.  \label{CW}
\end{equation}

Cnoidal solutions on the ring of radius $r_{0}$ must satisfy the
corresponding periodic boundary conditions (b.c.),
$\tilde{A}(y,t)=\tilde{A}(y+2\pi r_{0},t)$ (for the ordinary 1D
NLS equations with repulsion and attraction, analytical solutions
satisfying the periodic b.c. are collected in Refs.
\cite{Carr-repulsive} and \cite{Carr-attractive}, respectively).
Accordingly, parameter $H$ in solutions (\ref{cn}),
(\ref{modulus}) is not continuous, but rather takes discrete
values selected by matching the b.c. to the periodicity of
$\mathrm{cn}^{2}$,
\begin{equation}
\frac{\pi r_{0}}{n}=\sqrt{\frac{6\left( 1+I\right) }{B_{0}\left(
B_{1}+\left\vert B_{2}\right\vert \right) }}K(q),  \label{matching}
\end{equation}
where $K$ is the complete elliptic integral, and $n$ is an arbitrary integer.

Note that Eq. (\ref{new}) features the Galilean invariance: if
$\tilde{A}(y,t)$ is a solution, then its counterpart in the form
of a solution moving at an arbitrary velocity $v$ is
\begin{equation}
\tilde{A}_{c}(y,t)=\tilde{A}(y-vt)\exp \left(
\frac{i}{4}vy-\frac{i}{2}v^{2}t\right) ,  \label{Galileo}
\end{equation}
hence one can boost solitons (\ref{soliton}) and cnoidal waves (\ref{cn}) to
a (formally) arbitrary velocity by means of transformation (\ref{Galileo}).
In fact, the velocity is not arbitrary because the factor $\exp \left(
ivy/4\right) $ also must satisfy the periodic b.c., which leads to a
quantization condition, $v=4n/r_{0}$, with $n=0,\pm 1,\pm 2,\,...$ .

\subsection{Modulational instability of the uniform ring soliton in the
model with attraction}

In the model with attraction ($\chi =+1$), the existence of
uniform ring-shaped solitons trapped in annular troughs of large
radii is obvious, while a nontrivial issue is a possibility of
finding a stability region for such solutions. In terms of Eq.
(\ref{new}), the ring soliton is represented by a CW
(continuous-wave) solution with a constant amplitude $B_{0}$,
which was mentioned above as a limit case of the cnoidal solution
family, with $\mu =-2B_{0}^{2}$.

It is well known that, while all CW solutions of NLS-like equations with
attraction are modulationally unstable [including nonpolynomial equations
similar to Eq. (\ref{new}) \cite{Lennart}], periodic b.c. may stabilize them
if the CW amplitude is smaller than a certain critical value. This argument
suggests a possibility to find a stability region for the uniform ring
solitons in the present model with attraction.

A standard analysis of the MI assumes a perturbation of the CW solution,
\[
\tilde{A}(y,t)=\left[ B_{0}+B_{1}(y,t)\right] \exp \left( 2iB_{0}^{2}t+i\chi
_{1}(y,t)\right) ,
\]
with infinitesimal amplitude and phase perturbations $B_{1}$ and $\chi _{1}$.
Eigenmodes of the perturbations may be looked for assuming that they are
proportional to $\exp \left( ipy+\gamma t\right) $, where $p$ is an
arbitrary real wavenumber, and $\gamma $ is the corresponding instability
growth rate. Straightforward calculations demonstrate that the CW solution
to Eq. (\ref{new}) is subject to the MI under the condition
\begin{equation}
p^{2}\left[ 2B_{0}^{2}-(1+I)p^{2}\right] >0.  \label{MI}
\end{equation}

In the annular system that we are dealing with, $p$ is subjected to the
geometric quantization, the same way as above, i.e., $p=n/r_{0}$ with
integer $n$. Therefore, condition (\ref{MI}) with lowest $|n|=1$ is not
satisfied, making the CW solution \emph{\ modulationally stable}, under the
condition
\begin{equation}
2B_{0}^{2}r_{0}^{2}\leq 1+I.  \label{noMI}
\end{equation}
In a final form, this stability condition may be expressed in terms of the
full norm of the axisymmetric ring soliton. Substituting ansatz
(\ref{ansatz1}) in Eq. (\ref{Norm}), and making use of the
assumptions adopted above, one obtains $N\approx 8\pi r_{0}\rho
B_{0}^{2}$. Further, substituting here the above approximation,
$\rho =1/B_{0}$, yields
\begin{equation}
N=8\pi r_{0}B_{0}. \label{Nfinal}
\end{equation}
With regard to this relation, stability condition (\ref{noMI})
amounts to a limitation on $N$:
\begin{equation}
N<N_{\mathrm{thr}}=4\sqrt{2(1+I)}\pi \approx 22.55\text{.}  \label{thr}
\end{equation}
(note that this result does not contain any parameter).

A natural way to look at these stability conditions is to consider
a situation with gradually increasing $N$ (or $B_{0}$). The
azimuthally uniform ring solution will loose its stability when
the amplitude (or norm) attains the critical value defined by Eq.
(\ref{noMI}) [or by Eq. (\ref{thr})]. One may expect that the
onset of the MI gives rise to a \textit{bifurcation}, which
creates stable azimuthally modulated solutions, with a modulation
depth scaling as $\sqrt{N-N_{\mathrm{thr}}}$ for $0<\left(
N-N_{\mathrm{thr}}\right) /N_{\mathrm{thr}}\ll 1$. Alternatively,
one may fix the CW amplitude, $B_{0}$, and gradually increase the
trough's radius $r_{0}$; the loss of the CW stability and
emergence of the modulated solutions should occur when $r_{0}$
attains a critical value following from Eq. (\ref{noMI}),
\begin{equation}
\left( r_{0}\right) _{\mathrm{cr}}=\sqrt{(1+I)/2}B_{0}^{-1}.  \label{cr}
\end{equation}

In fact, the modulated states generated by the MI onset should be nothing
else but the stationary cnoidal wave (\ref{cn}), (\ref{modulus}). Indeed,
with regard to Eq. (\ref{CW}) and the fact that $K(0)=\pi /2$, it follows
form matching condition (\ref{matching}) that, with the increase of $r_{0}$
for fixed $B_{0}$, the cnoidal-wave solution appears, with an infinitely
small modulation depth, at $r_{0}=\left( r_{0}\right) _{\mathrm{cr}}$, where
the critical value $\left( r_{0}\right) _{\mathrm{cr}}$ is \emph{precisely
the same} as defined by the onset of the MI [i.e., given by Eq. (\ref{cr})].

A defect of the above approximate analysis is that, with regard to
relation $A_{0}=1/\rho $, stability condition (\ref{thr}) may be
cast in the form of $\rho /r_{0}>\sqrt{2/(1+I)}\approx \allowbreak
1.\allowbreak 1$, which \emph{does not} comply with the underlying
low-curvature assumption, $\rho _{0}\ll r_{0}$. This fact makes
the existence of modulationally stable ring-shaped solitons in the
present model with attraction doubtful. In direct simulations of
Eq. (\ref{gpe2D}) with $\chi =+1$, we were unable to find
modulationally stable annular uniform nonlinear states. In this
connection, it is relevant to recall that the linear limit of Eq.
(\ref{gpe2D}) with potential $V(r)=\varepsilon \cos (2r)$ does not
allow any stationary solution localized in the radial direction,
therefore the weak nonlinearity is essential in stabilizing the
axisymmetric rings.

Note that localized ring-shaped states may be possible in the
linear limit of the Bessel-lattice models considered in Refs.
\cite{kartashov} and \cite{repulsion}, with potentials
$-\varepsilon J_{0}\left( kr\right) $ and $-\varepsilon
J_{1}^{2}\left( kr\right) $, respectively; however, the azimuthal
(in)stability of such states was not studied in the presence of
attraction. On the other hand, \emph{stable} annular patterns
(both static ones and persistent ring-shaped breathers) were
recently found in an attractive model with a single circular
trough, which was created not by the linear potential, but rather
by modulation of the nonlinear coefficient in the radial direction
\cite{HS}. In that model, annular states may only be stable if the
annulus is \emph{wide enough}, with the ratio of its outer and
inner radii exceeding $0.47$. Vorticity-carrying rings were found
to be unstable in the same setting.

\subsection{Gap Soliton nature of ring shaped states}

In this section we investigate the gap soliton nature and the band
structure associated with ring shaped states in terms of the
following  2D GPE
\begin{equation}
i \psi_t=-\Delta_r \psi + \varepsilon \cos(2 r) \psi + \chi
|\psi|^2 \psi \label{GPEspher}
\end{equation}
where $\Delta_r$ denotes the spherical 2D laplacian:
$\Delta_r=\frac{\partial^2}{\partial r^2} + \frac{2}{r} \frac
{\partial}{\partial r} - \frac{\hat l^2}{r^2}$, with $\hat
l^2=-\frac{\partial^2}{\partial \varphi^2}$ the squared 2D angular
momentum. This equation corresponds to the $\theta\rightarrow
\frac \pi 2$ limit  ($u(r,\pi/2,\varphi)\rightarrow
\psi(r,\varphi)$) of the spherically symmetric 3D GPE, with
$(r,\theta,\varphi)$ denoting spherical coordinates. Notice that
apart for the factor of two in front of the term $\frac 1 r
\frac{\partial }{\partial r}$, due to spherical rather than
cylindrical symmetry, Eq. (\ref{GPEspher}) is the same as Eq.
(\ref{gpe2D}). From a physical point of view Eq. (\ref{GPEspher})
is appropriate to describe a condensate trapped in the outer part
of the conical region created by a cylindrical beam of light
passed through a very short focal distance lens. In the following
we shall refer to this setting as the "pizza" geometry since, in
contrast with the pancake geometry, the thickness of the
condensate in the axial direction is not uniform and  reduces to
zero as $r \rightarrow 0$. This geometry could be implemented by
combining a cylindrical beam creating the radial lattice with a
blue-detuned conical beam (the pizza-shaper)
\begin{figure}[htb]
\centerline{
\includegraphics[width=4cm,height=4cm,clip]{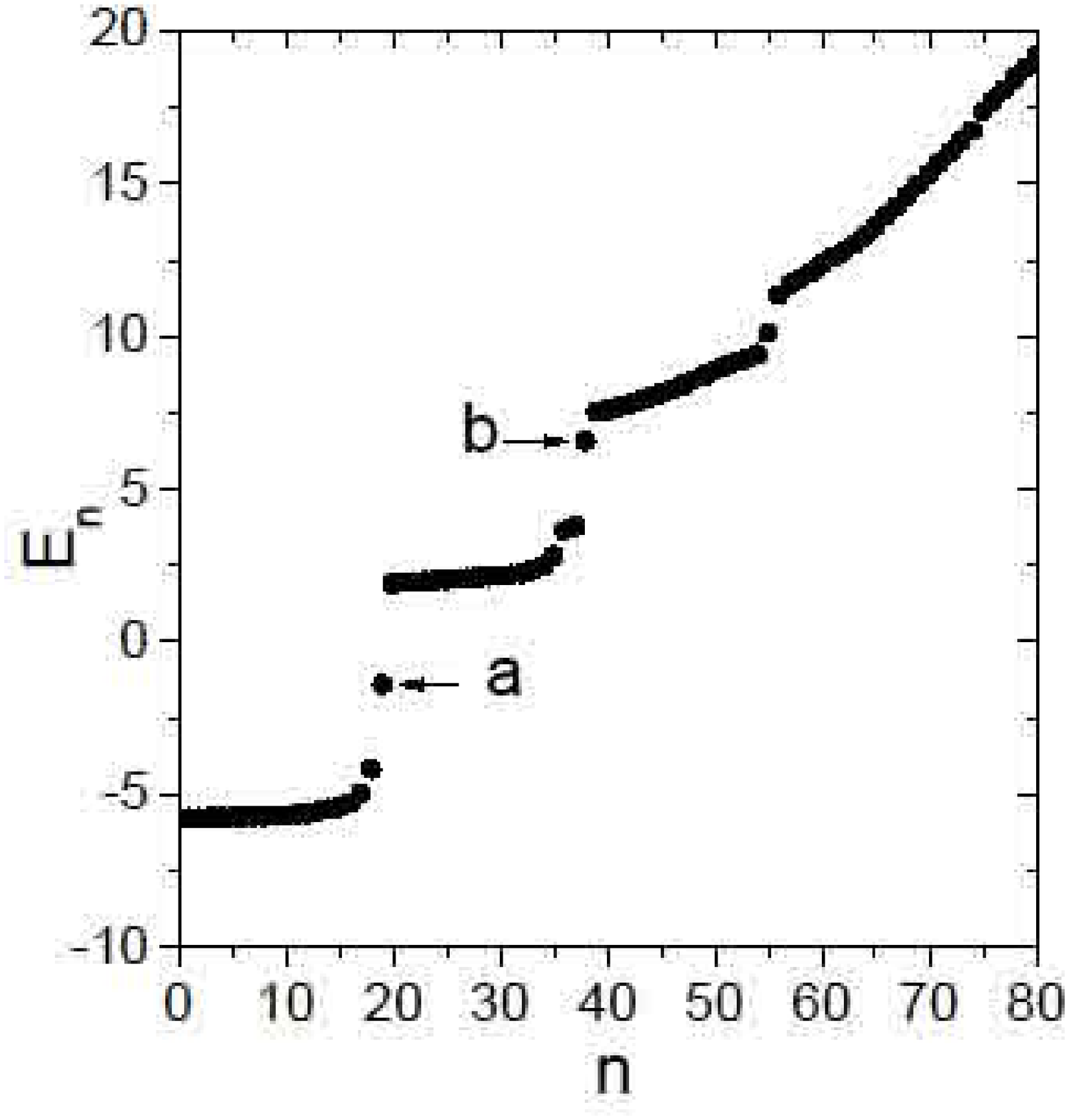}
\includegraphics[width=4cm,height=4cm,clip]{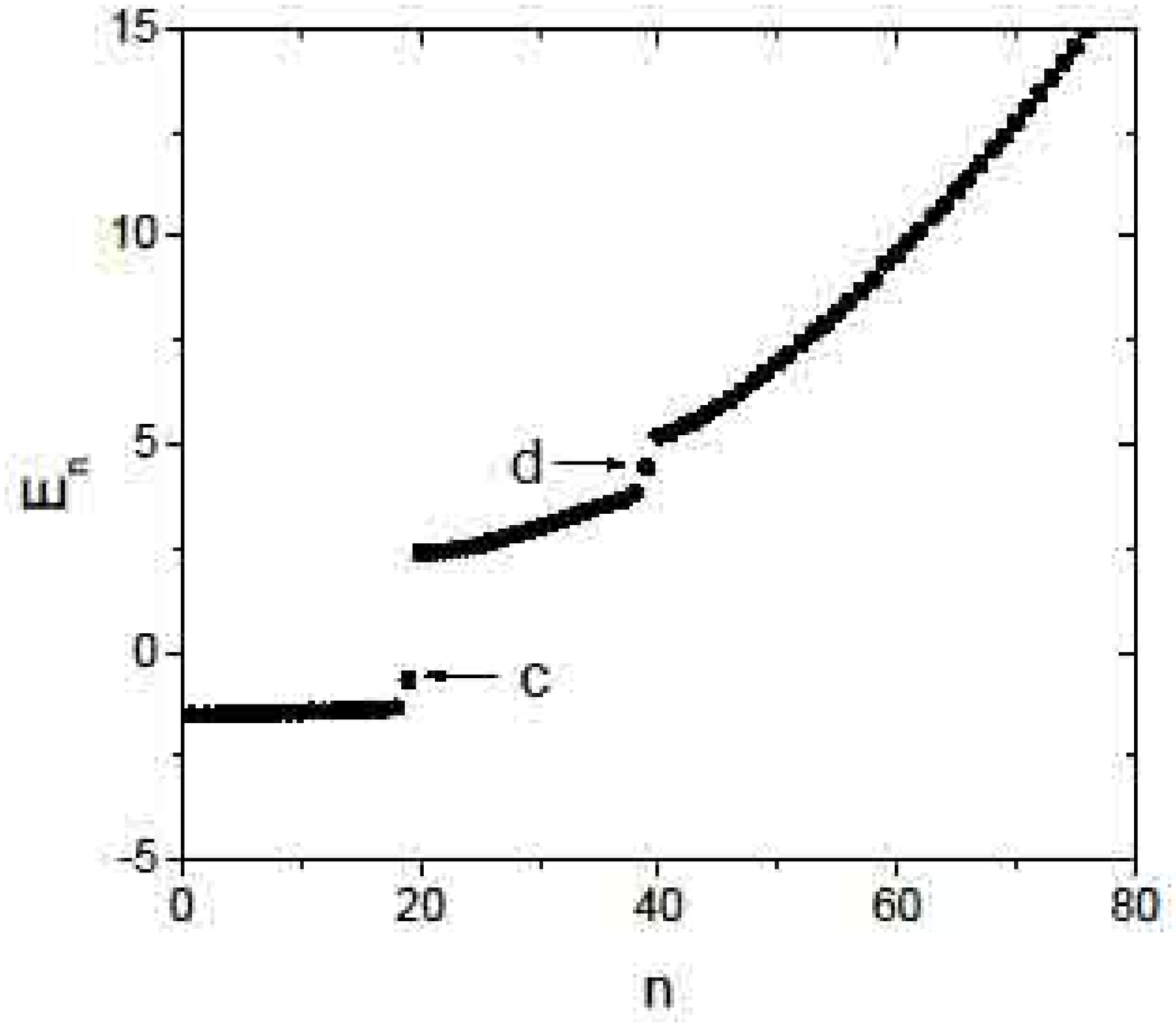}}
\caption{Left Panel: Energy levels of Eq. (\ref{reducedradial})
with $\chi=-1$, $l=10$ and $\varepsilon=10$. Right panel: The same
as in the left panel for the repulsive case $\chi=1$ with $l=2$
and $\varepsilon=-4$.} \label{fig8}
\end{figure}
launched in the opposite direction. Notice that at large radial
distances (i.e. in the outer part of the pizza) the differences
between pancake and pizza geometry become negligible especially if
the thickness of the border is very small compared to the radius.
Ring shaped states of Eq. (\ref{GPEspher}) located in potential
wells far from the center are expected, therefore, to be
qualitatively similar to the ones of the pancake setting Eq.
(\ref{gpe2D}).

In the following we look for ring gap solitons of Eq.
(\ref{GPEspher}) in the form  of stationary states
$\psi(r,\varphi,t)$ of the form
\begin{equation}
\psi(r,\varphi,t)=\phi(r)e^{-i\mu t} e^{i l \varphi}. \label{stat}
\end{equation}
By substituting Eq. (\ref{stat}) into Eq.(\ref{GPEspher}) we
obtain the following equation for the radial function $\phi(r)$
\begin{equation}
\phi_{rr}+\frac{2}{r}\phi_r+ (\mu-\varepsilon \cos(2 r)) \phi
-(\chi |\phi|^2 + \frac{l^2}{r^2})\phi =0, \label{radial}
\end{equation}
where  $l$ are integer numbers which allow to satisfy the periodic
boundary condition in $\varphi$:
$\psi(r,\varphi+2\pi,t)=\psi(r,\varphi,t)$.  Equation
(\ref{radial}) can be mapped into a 1D eigenvalue problem by means
of the transformation $\phi(r)= \frac {\xi(r)}{r} $ leading to the
following reduced radial equation
\begin{equation}
\xi_{rr} +  [E-\varepsilon \cos(2 r) -\frac { l^2 +
\chi|\xi|^2}{r^2}]\xi =0 \label{reducedradial}
\end{equation}
which has the form of a 1D Schr\"odinger equation with periodic
potential and nonlinear effective centrifugal barrier: $(l^2 +
\chi|\xi|^2)/r^2$ (here $E\equiv \mu$). For sufficiently large
distances $r \gg 1$ and fixed angular momentum and nonlinearity,
this equation reduces to the Mathieu equation with the well known
band gap structure. The part of the spectrum associated to
excitations localized nearby the origin, however, is expected to
be strongly influenced by the centrifugal barrier which at short
distances enhances the nonlinearity and favors the formation of
localized states.

To investigate the band structure and the gap solitons of Eq.
(\ref{reducedradial}) we use a self-consistent method developed in
Ref. \cite{ms05}. The gap solitons obtained with the
self-consistent method will then be used as initial conditions for
time propagation of the radial 2D GPE in Eq. (\ref{gpe2D}).
\begin{figure}[htb]
\centerline{
\includegraphics[width=3.5cm,height=3.5cm,clip]{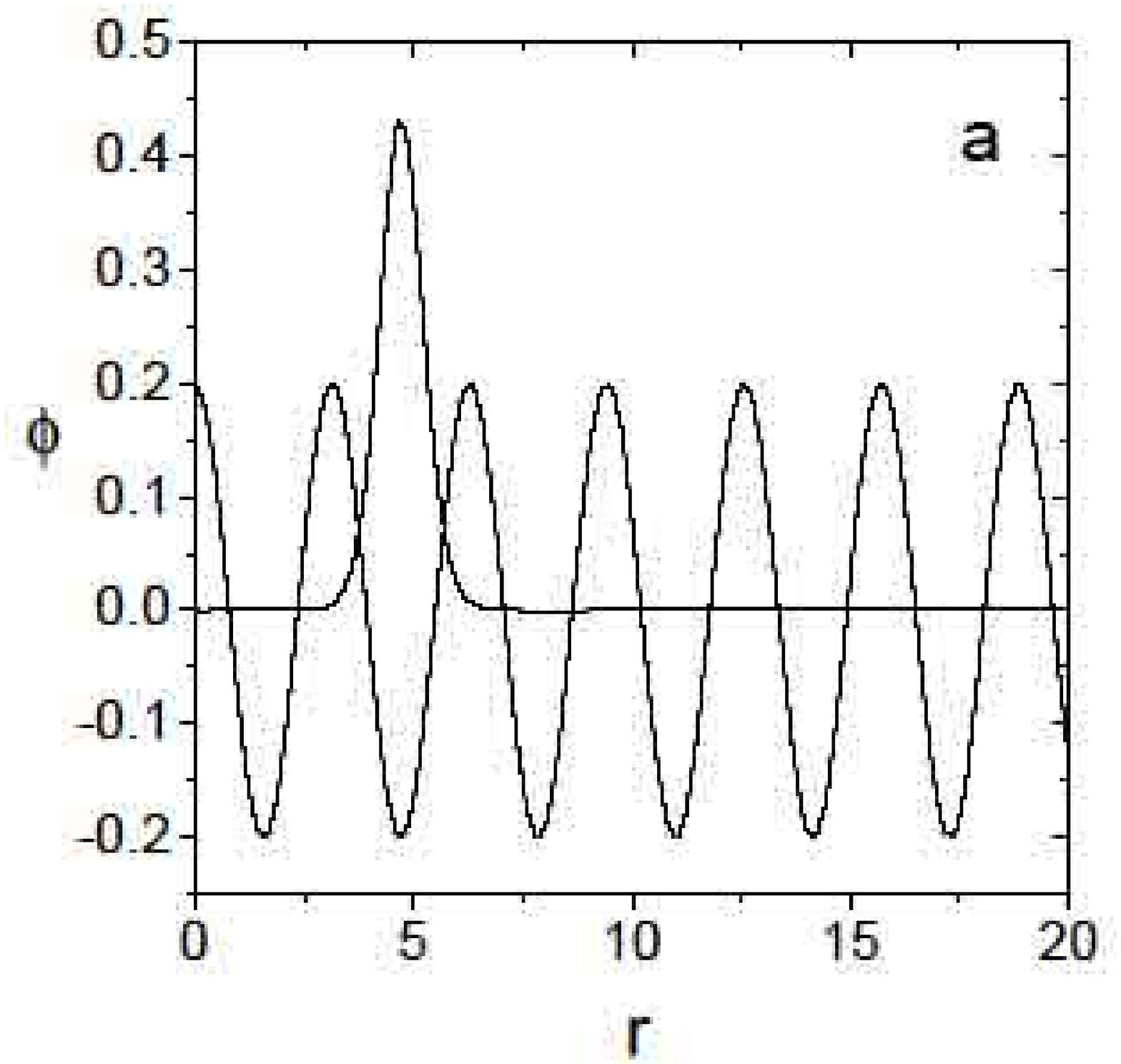}
\includegraphics[width=3.5cm,height=3.4cm,clip]{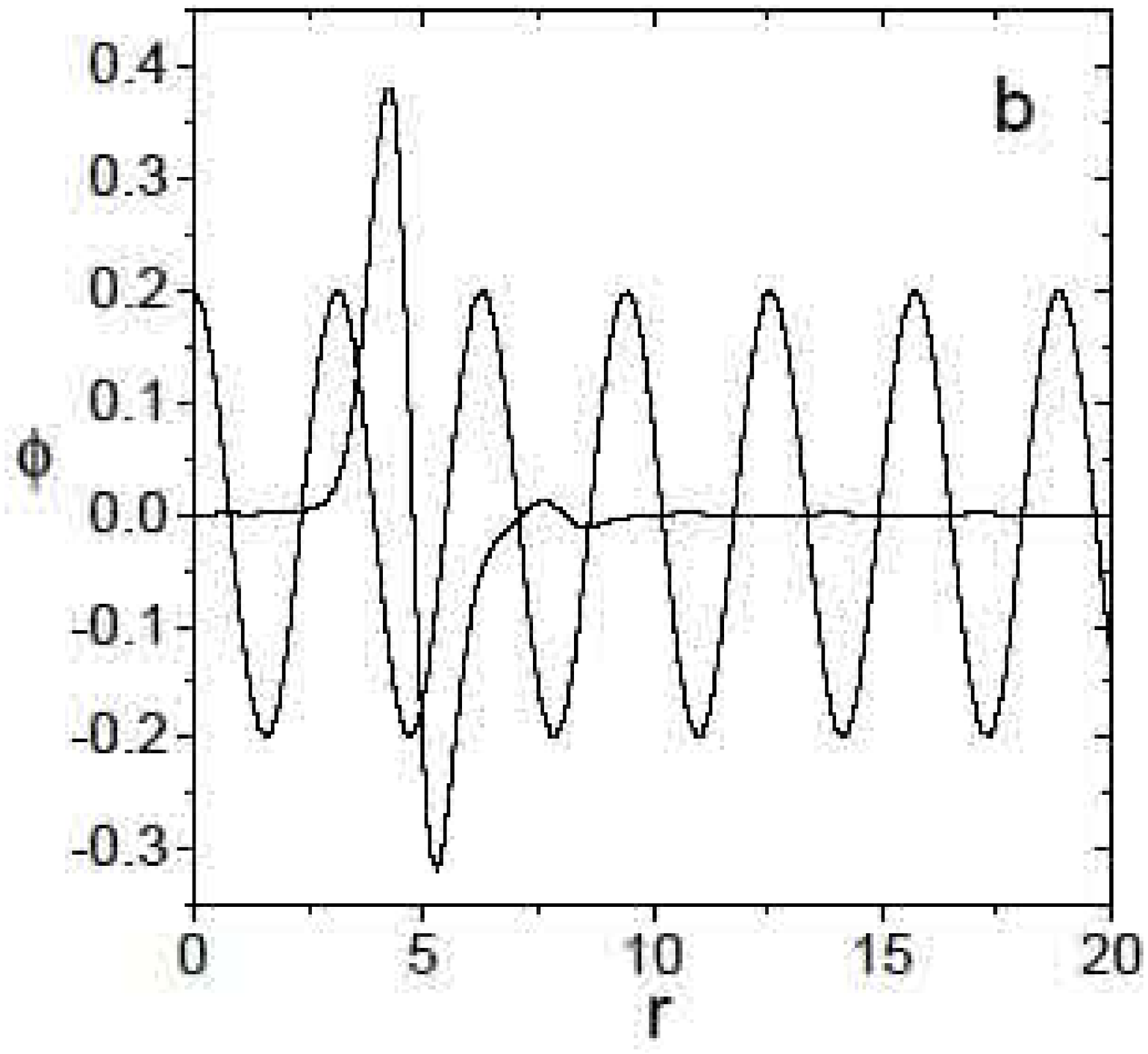}}
\centerline{
\includegraphics[width=3.5cm,height=3.5cm,clip]{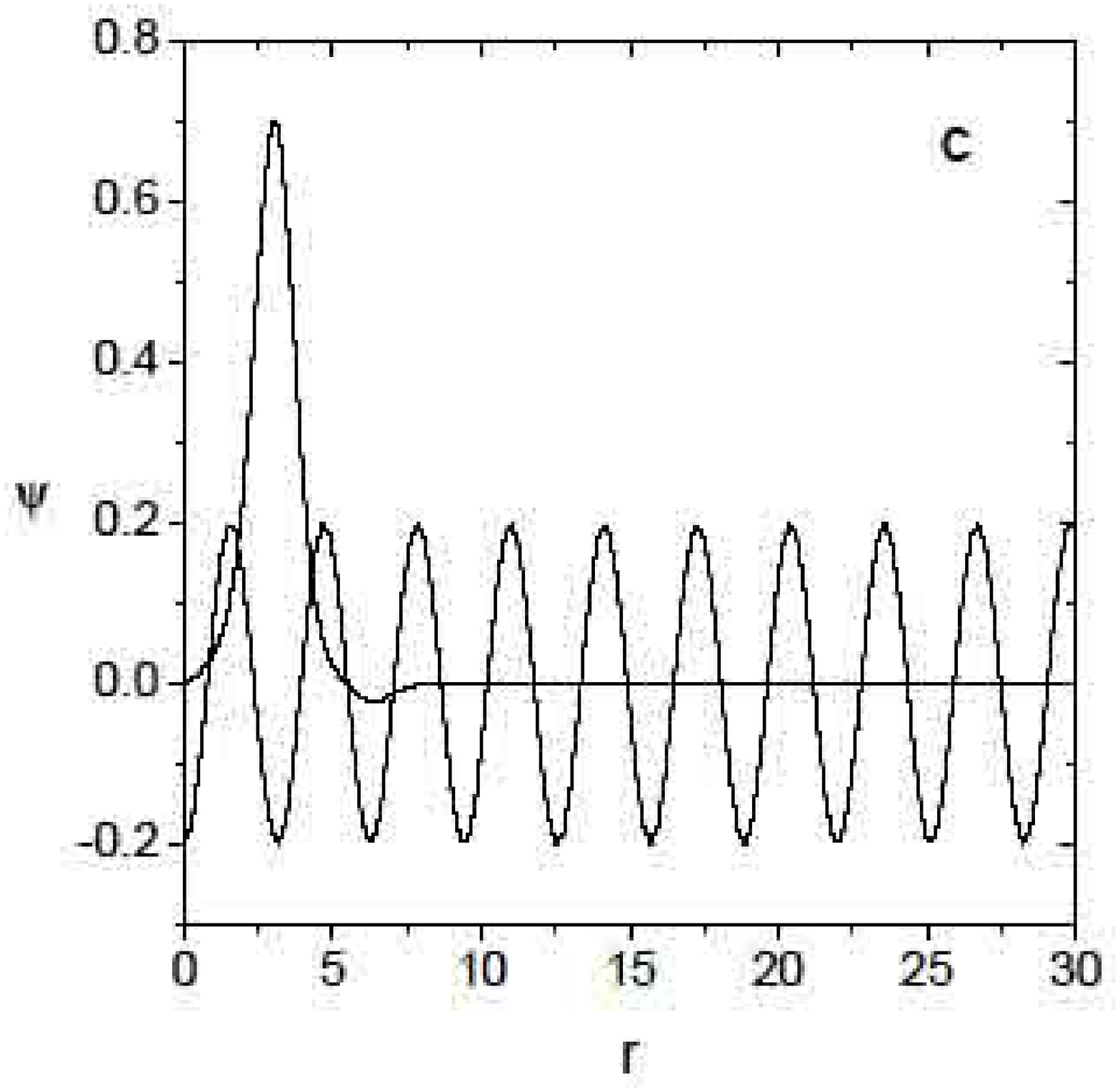}
\includegraphics[width=3.5cm,height=3.5cm,clip]{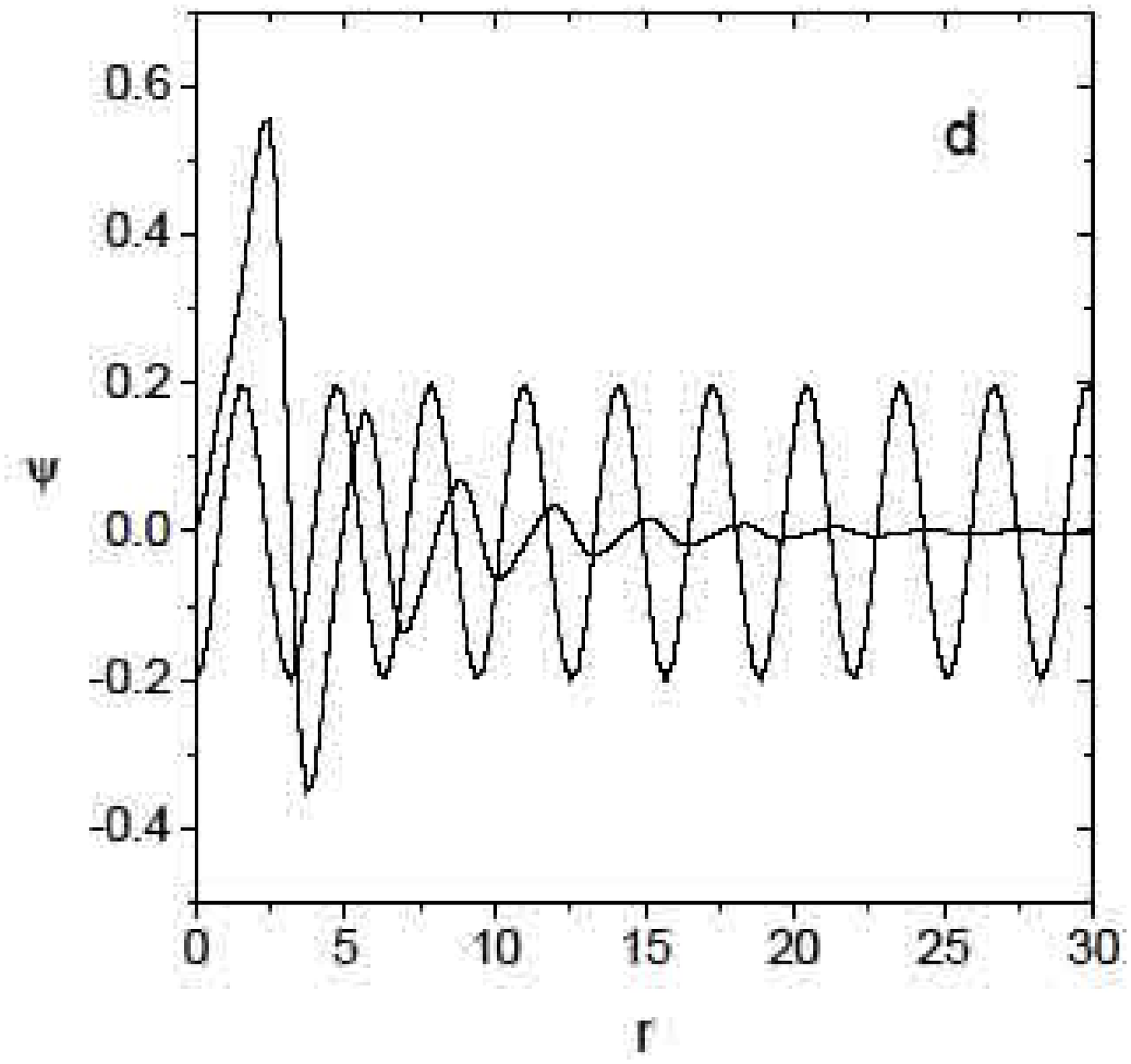}
} \caption{Panels a,b. Attractive radial gap solitons (thick
lines) trapped in circular troughs of the potential at $r=1.5\pi$
and corresponding to the energy levels $a,b$ of the left panel of
Fig. \ref{fig8}, respectively. Panels c,d. Repulsive radial gap
solitons (thick lines) trapped in circular troughs of the
potential at $r=1.5\pi$ and corresponding to the energy levels
$c,d$ of the right panel of Fig. \ref{fig8}, respectively. The
thin lines corresponds to the radial potential $\varepsilon \cos(2
r)$ scaled by a factor $50$ in the left panel and by a factor $20$
in the  right panel, for graphical convenience.} \label{fig9}
\end{figure}
\begin{figure}[htb]
\centerline{
\includegraphics[width=4cm,height=3cm,clip]{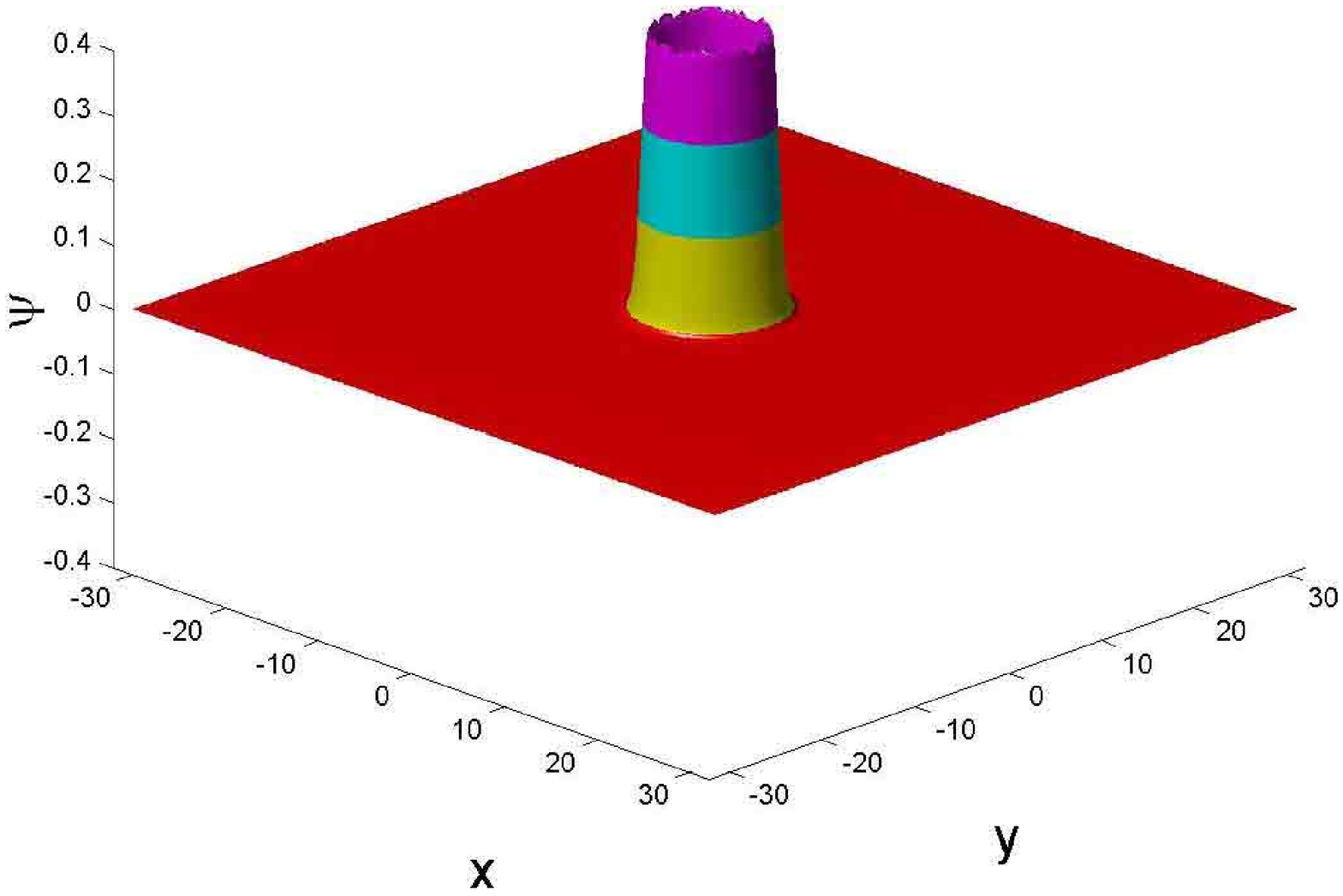}
\includegraphics[width=4cm,height=3cm,clip]{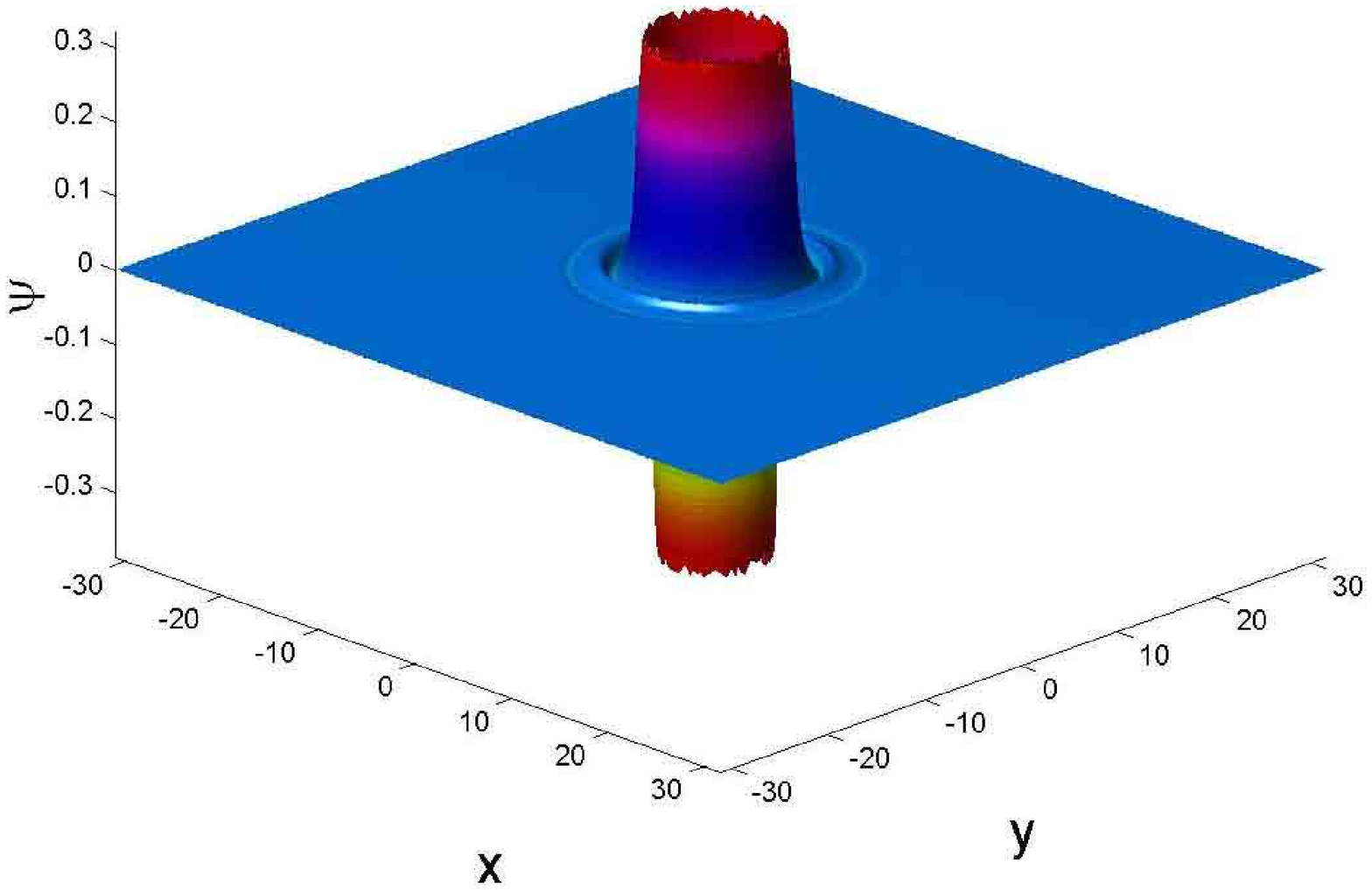}}
\centerline{
\includegraphics[width=4cm,height=3cm,clip]{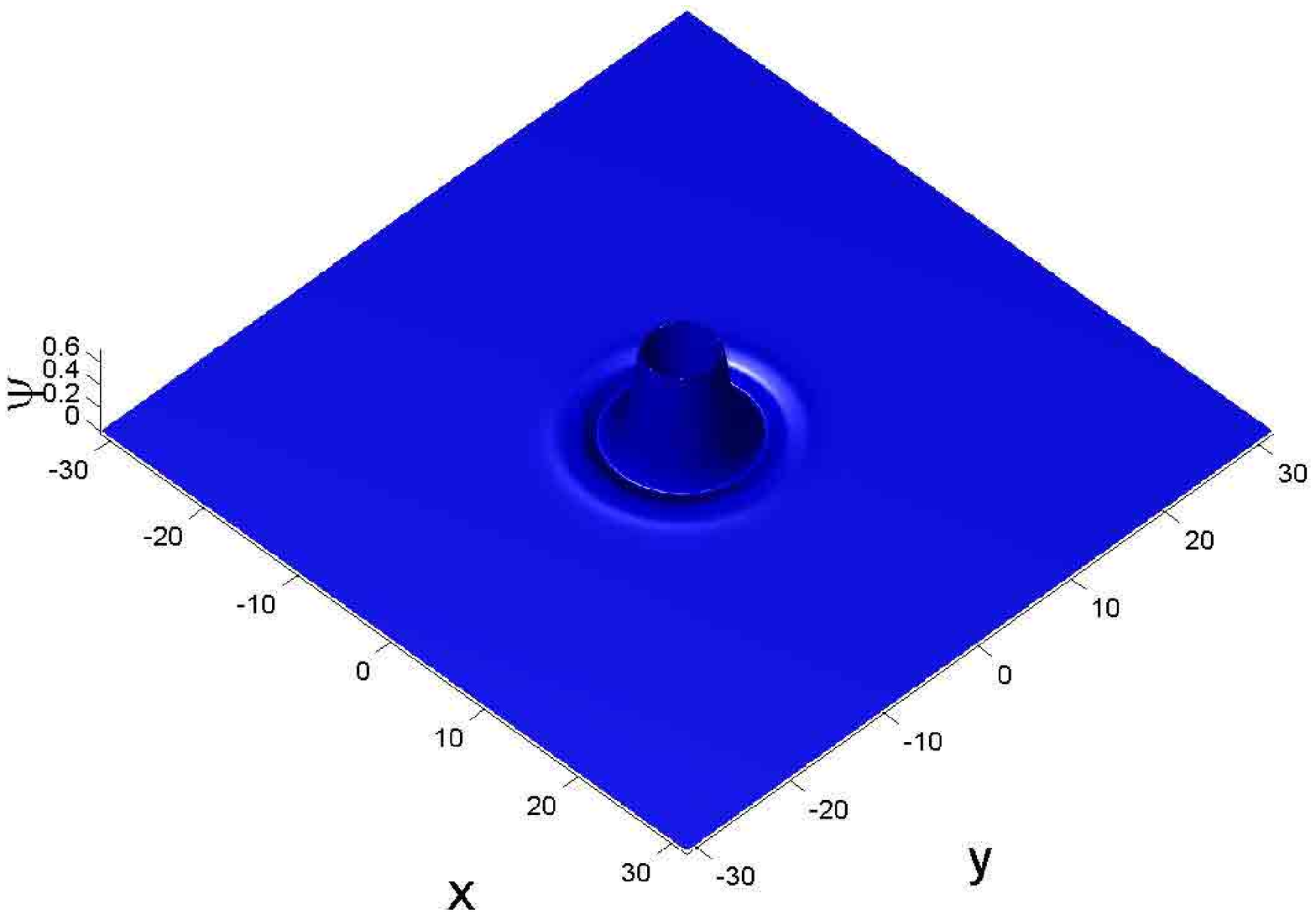}
\includegraphics[width=4cm,height=3cm,clip]{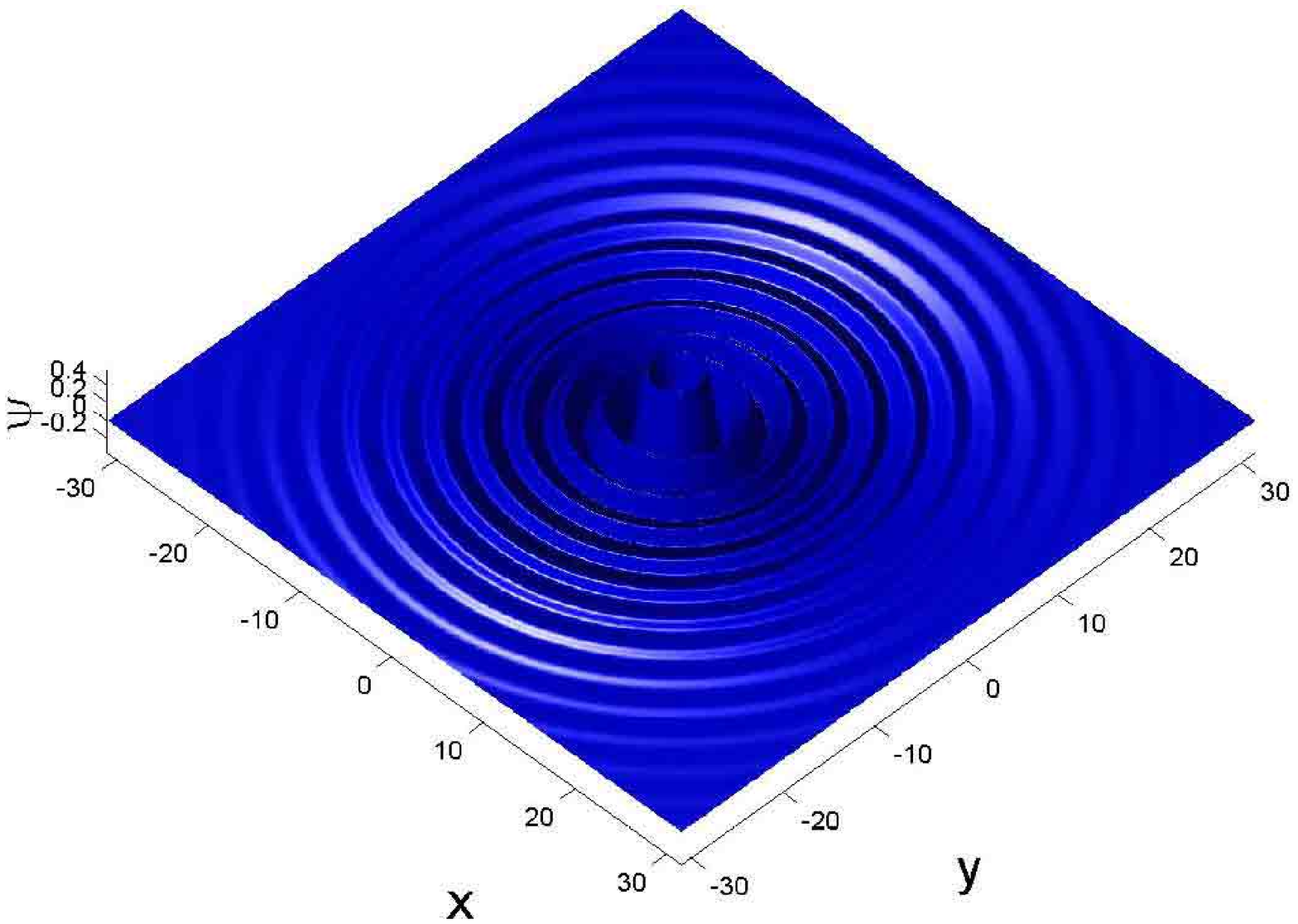}
} \caption{Top panels. 3D view of the gap solitons corresponding
to the bound states a (left panel) and b (right panel) of Fig.
\ref{fig8}. Bottom panels. 3D view of the gap solitons
corresponding to the bound states c (left panel) and d (right
panel) of Fig. \ref{fig8}.} \label{fig10}
\end{figure}
As a result we show the existence of radially symmetric gap
solitons which correspond to the ring shaped states discussed
before.
\begin{figure}[htb]
\centerline{\includegraphics[width=3.5cm,height=3.5cm,clip]{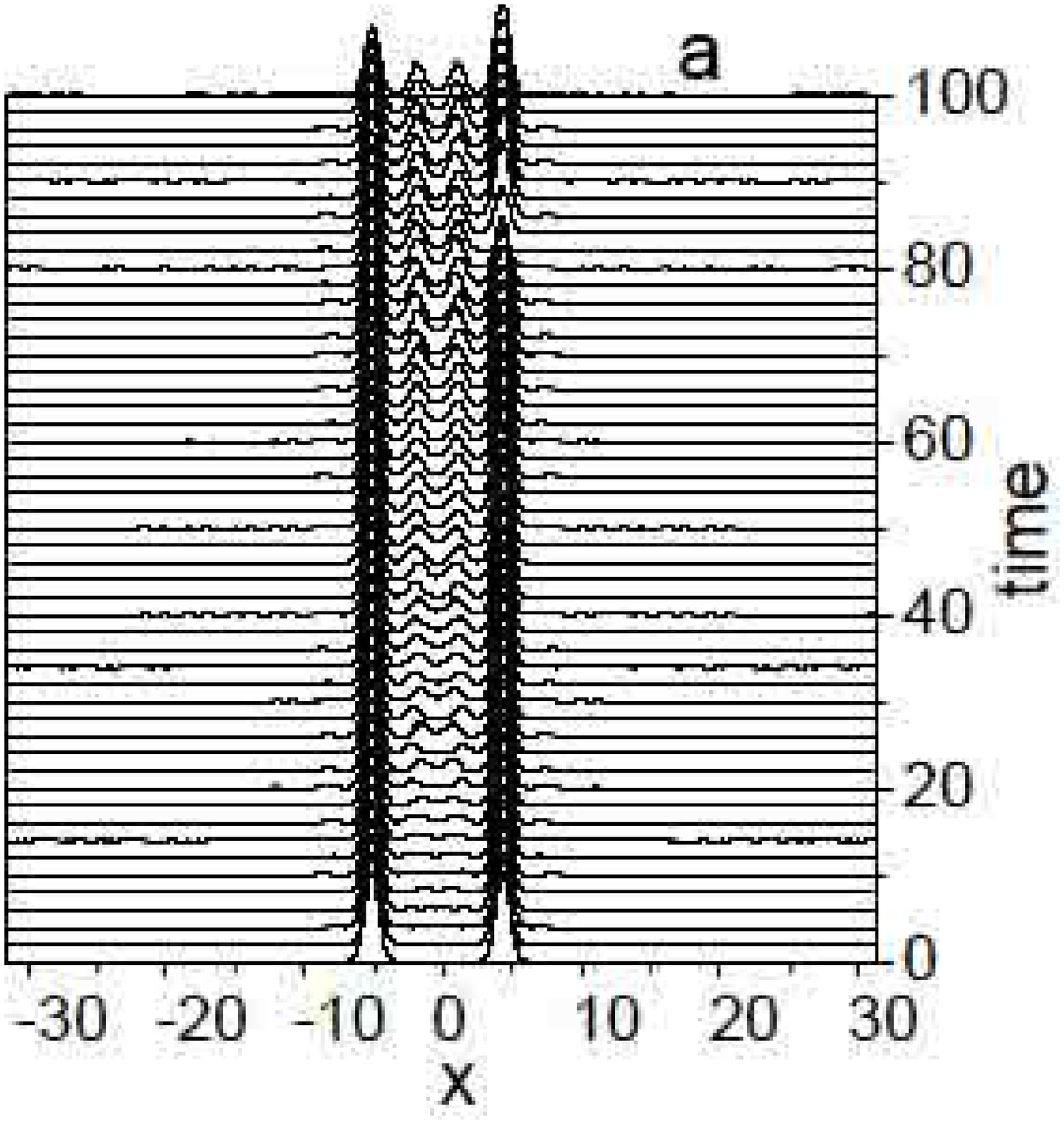}
\includegraphics[width=3.5cm,height=3.5cm,clip]{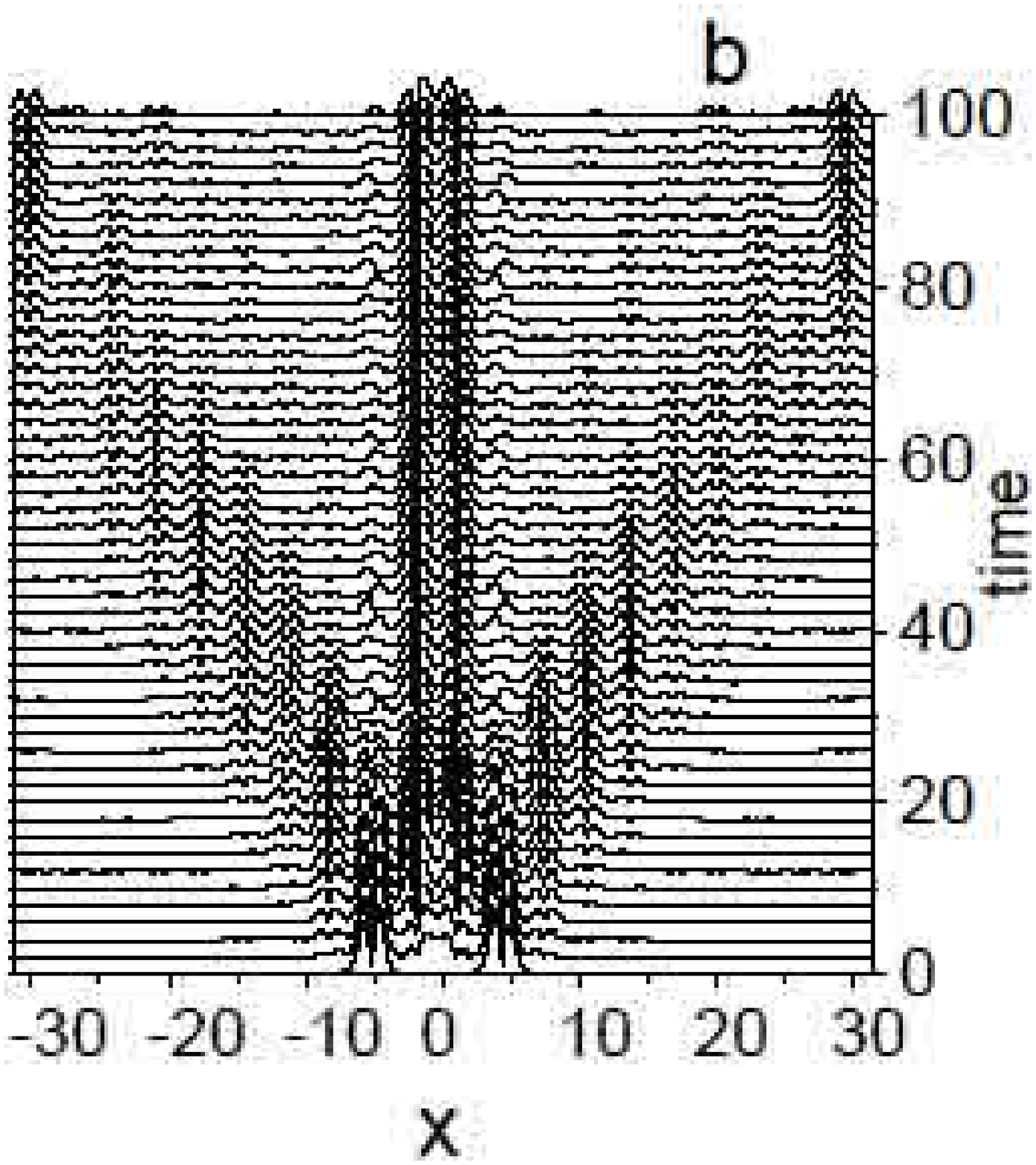}}
\centerline{\includegraphics[width=3.5cm,height=3.5cm,clip]{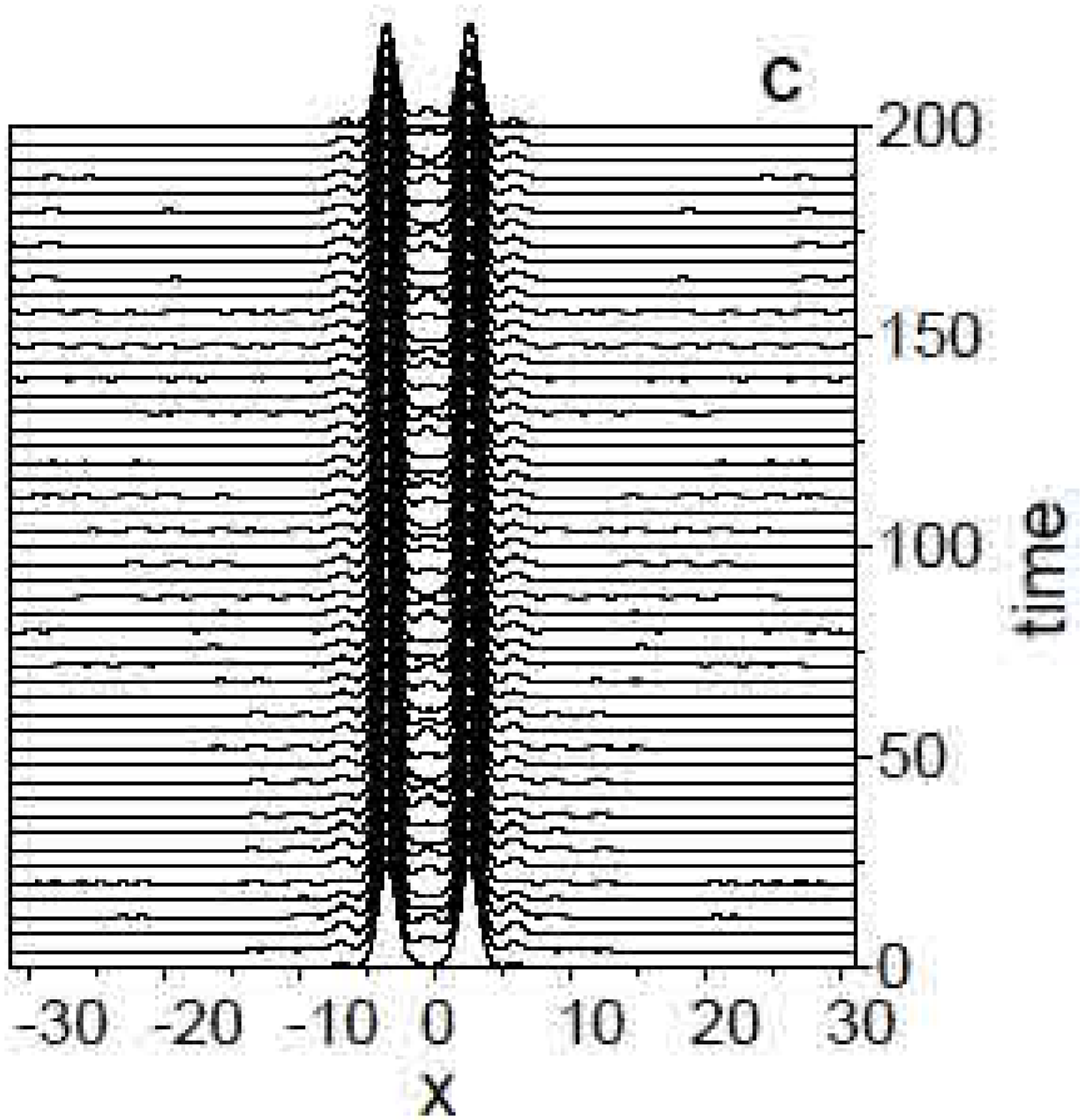}
\includegraphics[width=3.5cm,height=3.5cm,clip]{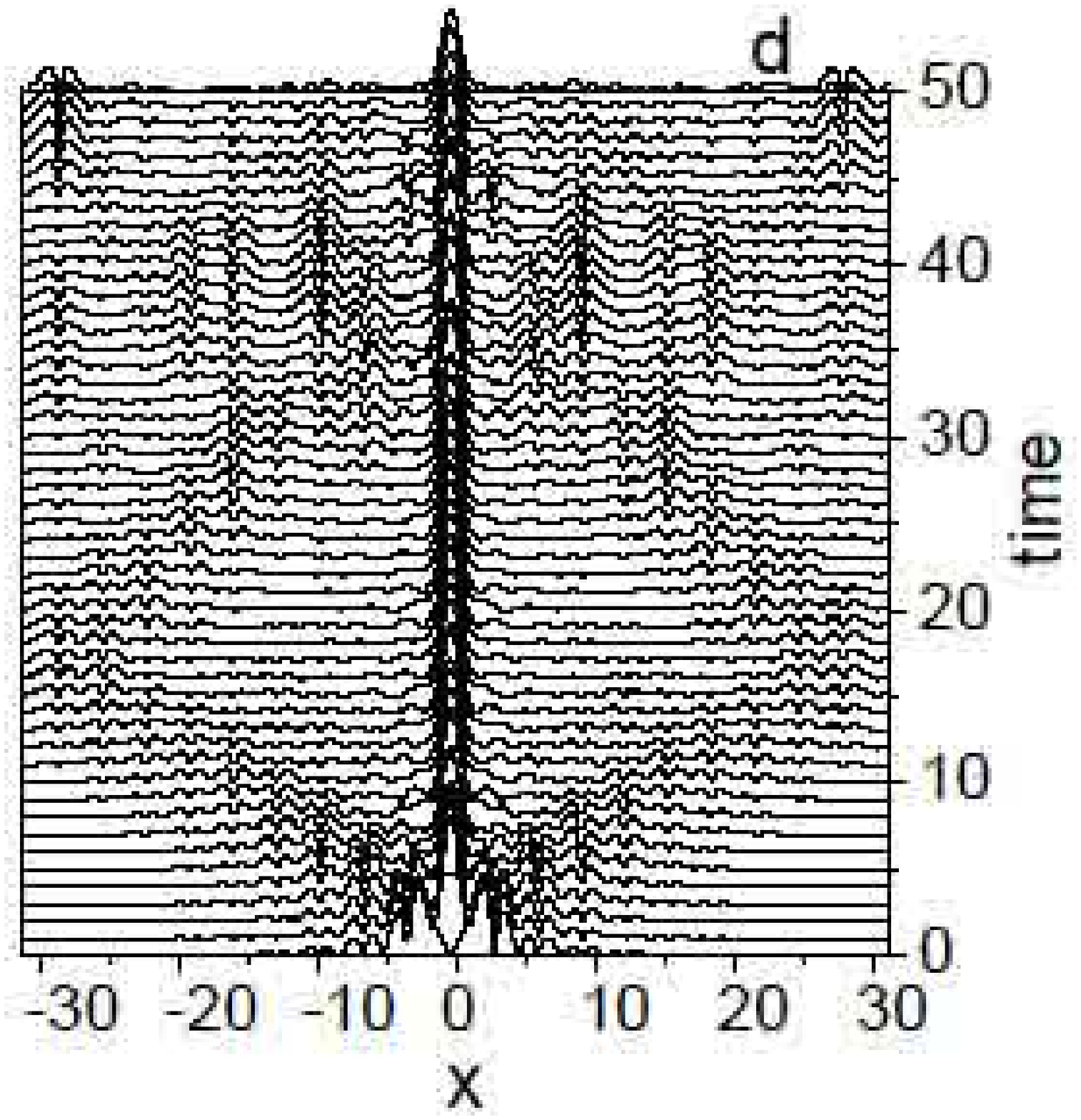}}
\caption{Time evolution ($y=0$) section of $|\psi|^2$ obtained
from Eq. (\ref{gpe2D}) by taking as initial conditions the
localized states depicted in panels a-d of Fig. \ref{fig9}. The
norm of the attractive  states in panels $a,b$ are $4.85$, $4.81$,
respectively, and correspond to bound states a,b, of Fig.
\ref{fig8}. Repulsive states in panels $c,d$ have norm  $11.47$
and $9.17$, respectively, and correspond to the bound states c,d,
of Fig. \ref{fig8}. Parameters are the same as in corresponding
panels of  Fig. \ref{fig8}. } \label{fig11}
\end{figure}
In Fig. \ref{fig8} we depict the lower energy part of the spectrum
obtained from Eq. (\ref{reducedradial}) in the case of attractive
and repulsive interactions. We see in both cases the existence of
a series of discrete levels located in the gaps between the bands
which correspond to radially symmetric localized states. The wave
functions of  the energy levels labeled a-d in Fig. \ref{fig8} are
shown in Fig. \ref{fig9}.

The symmetries of these localized states is similar to the ones of
intrinsic localized modes of nonlinear lattices and gap solitons
of continuous 1D GPE with a periodic potential. Adopting the same
terminology introduced for these cases we shall refer to them as
the onsite symmetric and the onsite asymmetric ring gap solitons.
A 3D plot of these states is reported in Fig. \ref{fig10} (notice
that the asymmetric state gives two rings of matter located in the
same trough of the potential in the attractive case and many ring
oscillations in the repulsive case ). The stability of these gap
solitons has been checked by time propagation under the 2D GPE in
Eq. (\ref{gpe2D}). The results are reported in Fig. \ref{fig11}
for the $y=0$ section of of $|\psi(x,y,t)|^2$ as a function of
time. We see that for the attractive case the gap solitons are
always unstable, the onsite symmetric state  becomes
modulationally unstable at $t\approx 80$ while the asymmetric
state decays into a state localized in the center and a series of
ring solitons which expand trough the lattice keeping the same
symmetry as the initial state.  For the repulsive case we see that
while the asymmetric state in the second band gap (panel d) is
unstable, the onsite symmetric ring gap soliton in the first band
gap is quite stable under time evolution. The stability of the
onsite symmetric  repulsive ring gap solitons and the modulational
instability of the repulsive ones has been found also for other
parameter values and seems to be  a general property of the model.
The existence of other types of ring gap solitons (which resemble
intersite symmetric and asymmetric intrinsic localized modes of
nonlinear lattices) is also possible but they look even more
unstable under time evolution than the onsite asymmetric ones for
both attractive and repulsive cases. In the following section we
shall investigate onsite symmetric gap soliton states in more
details.

\subsection{Numerical investigation of ring-shaped states}

\begin{figure}[tbh]
\centerline{
\includegraphics[width=4cm,height=3cm,clip]{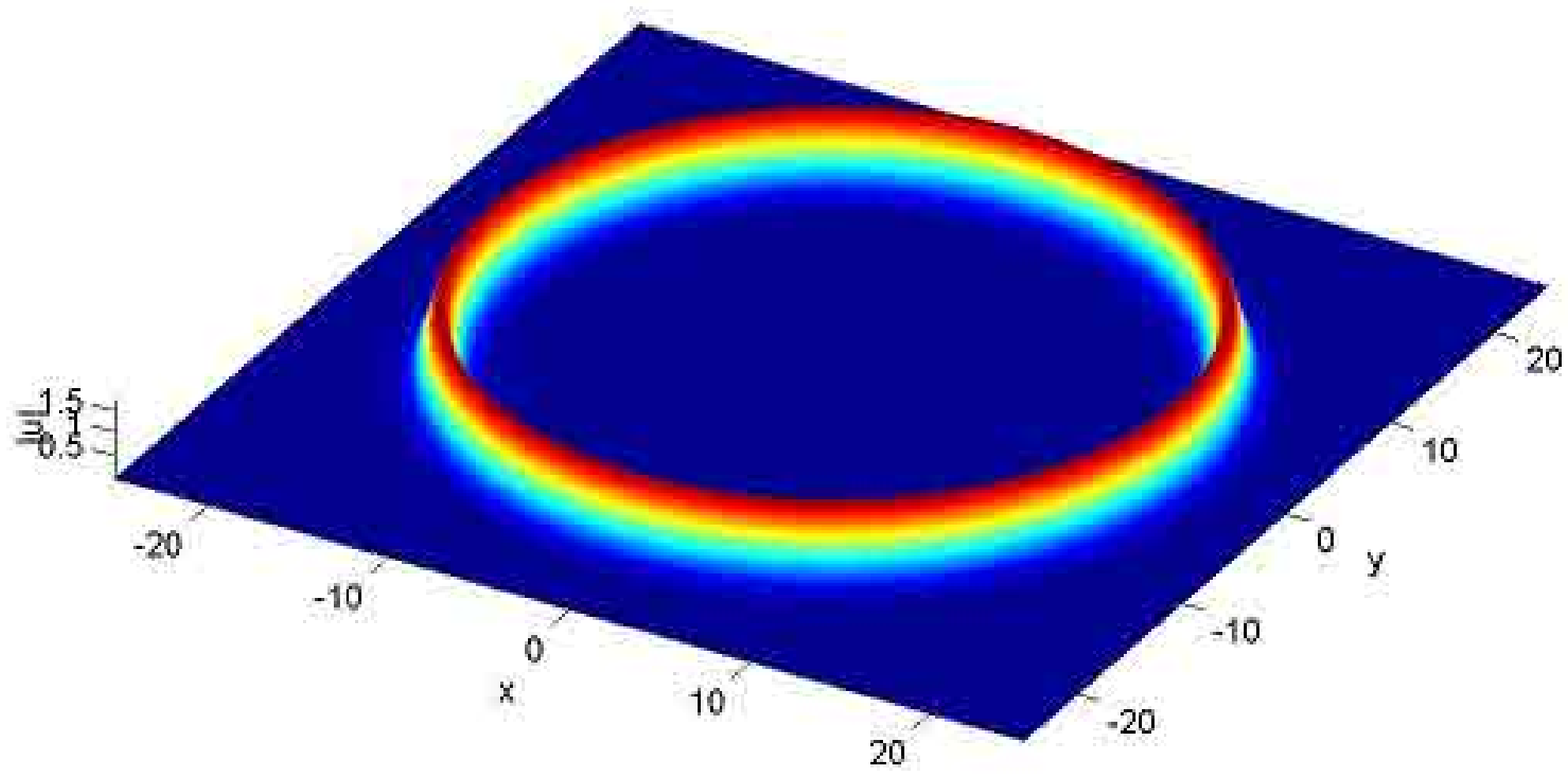} \quad
\includegraphics[width=4cm,height=3cm,clip]{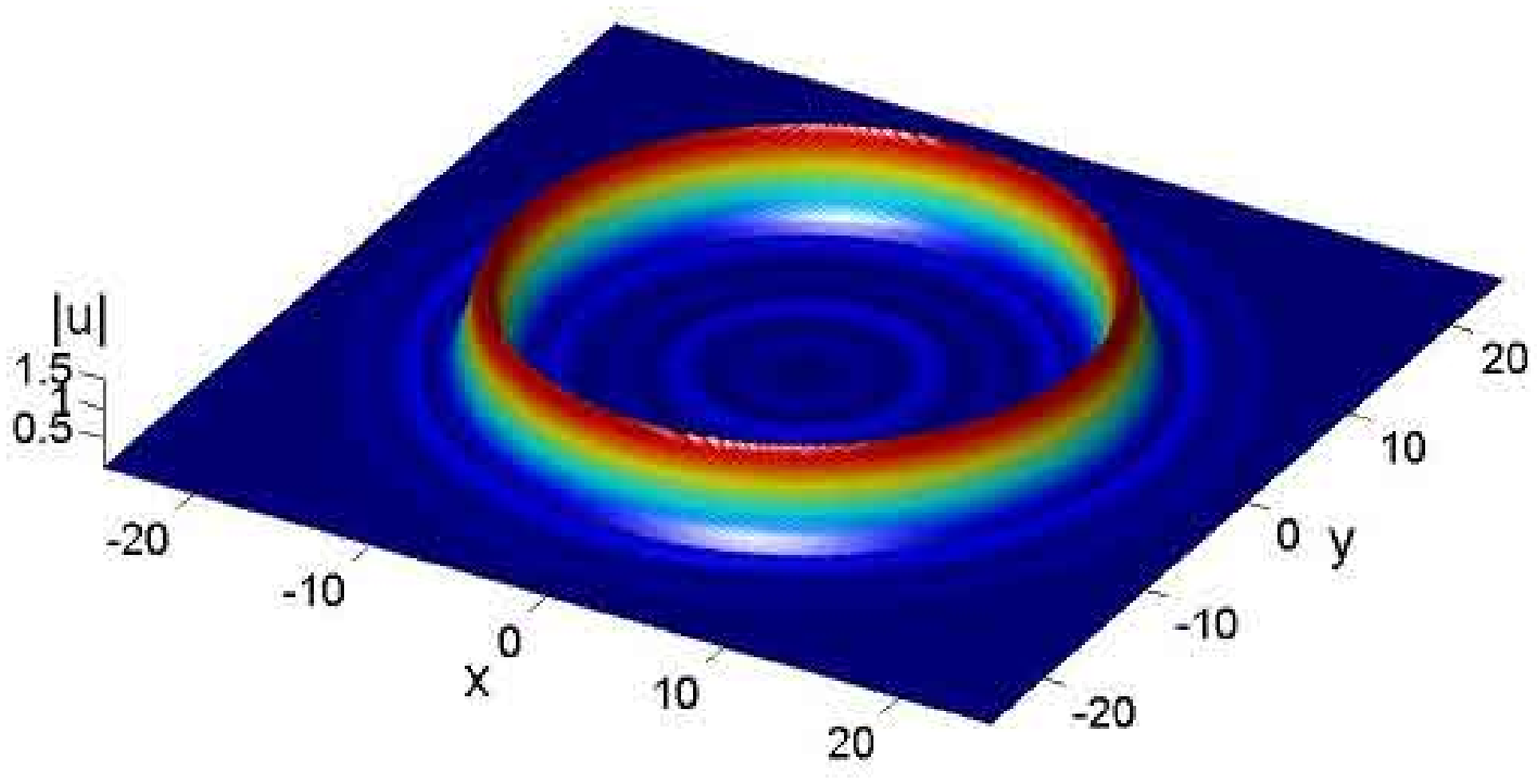}
} \caption{Left panel: An unstable ring-shaped soliton in the
attractive model with $\protect\varepsilon =2$, trapped in a
circular trough at $r_{0}=6\protect\pi $. The amplitude of this
soliton is $1.8$, and its chemical potential and norm are
$\protect\mu =-2.3$ and $N\simeq 420$. Note that analytical
approximation (\protect\ref{Nfinal}) yields, in this case,
$N\simeq 426$, which is very close to the numerically found value,
and its modulational instability is definitely predicted by Eq.
(\protect\ref{thr}). Right panel: A stable ring-shaped soliton
(actually, it is a \textit{radial gap soliton}) in the repulsive
model with $\protect\varepsilon =3$, trapped at
$r_{0}=5\protect\pi $. In the latter case, $N=344.3$ and
$\protect\mu =0.78$.} \label{fig12}
\end{figure}

Generic examples of a ring-shaped soliton in the attractive model,
and its counterpart in the repulsive one, found from direct
solution of GPE (\ref{gpe2D}) in, respectively, imaginary and real
time, are displayed in Fig. \ref{fig12} (in the latter case,
real-time simulations were run with absorbers set at $r=8\pi $ and
$r=0$). The latter example represents \textit{radial gap solitons}
with the core trapped in a remote circular trough (cf. the right
panel in Fig. \ref{fig4} that shows a radial gap soliton with the
core trapped in the center). The fact that the soliton belongs to
the gap type is clearly attested to by conspicuous inner and outer
radial fringes attached to the core (the fringes are absent in the
ordinary soliton in the left panel of the figure).

\begin{figure}[tbh]
\centerline{\includegraphics[width=8cm,height=6cm,clip]{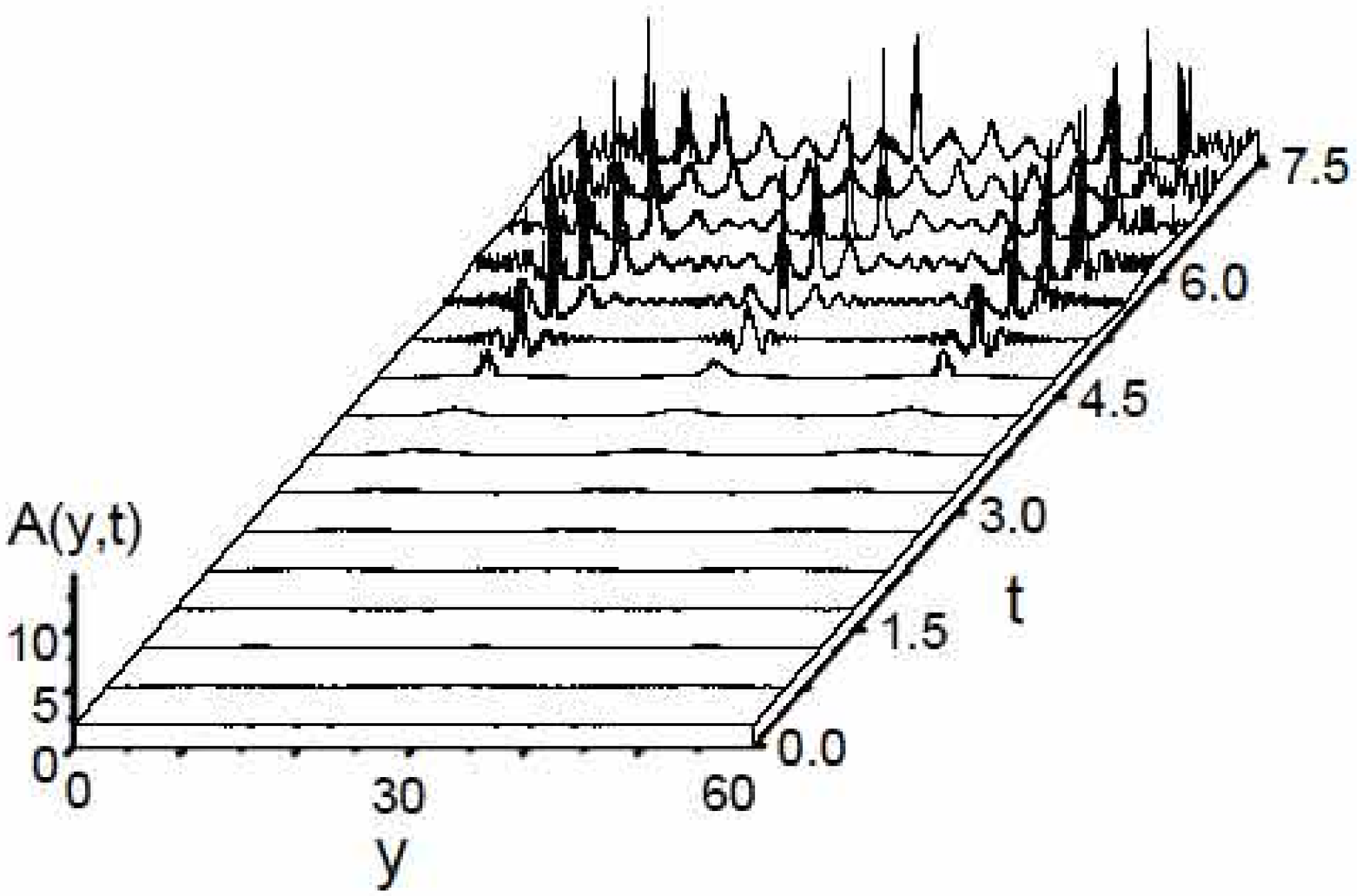}
{\bf (a)}}
\centerline{\includegraphics[width=8cm,height=6cm,clip]{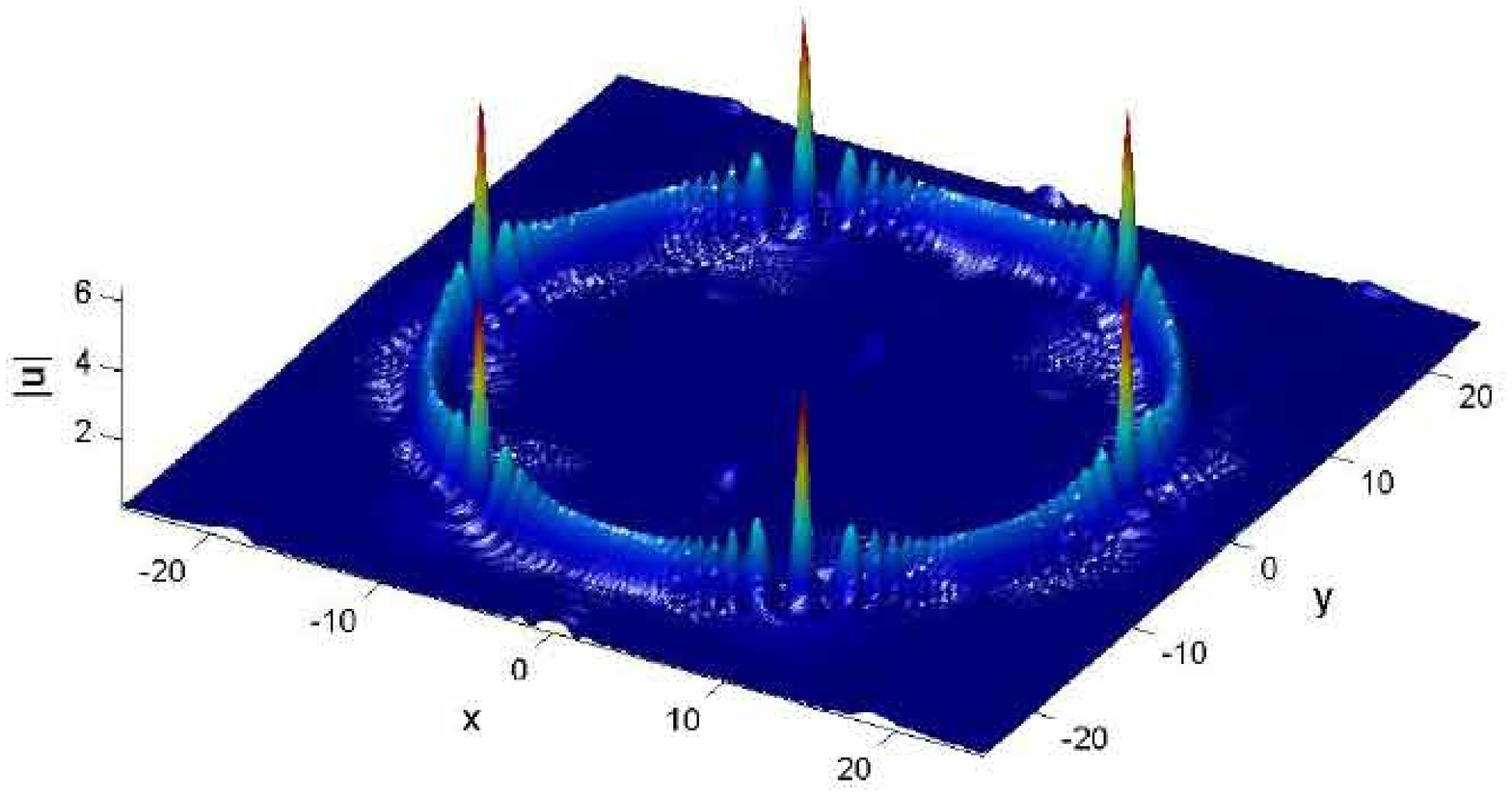}{\bf
(b)}} \caption{(a) Development of the modulational instability of
the axisymmetric state in Fig. \protect\ref{fig12} (left panel),
triggered by a small azimuthal perturbation, $\protect\delta
u=0.02\,\cos (6\protect\phi )\,\exp \left( -(r-r_{0})^{2}/2\right)
$. The evolution is shown in terms of the amplitude, $|A(y,t)|$,
taken along circumference $y=r_{0}\protect\theta $, with
$r_{0}=6\protect\pi $, $\protect\theta $ running from $0$ to
$\protect\pi $ (due to the symmetry, only half of the circle is
shown). (b) The field profile formed by $t=5$, demonstrating the
emergence of a regular necklace-shaped pattern.} \label{fig13}
\end{figure}

The evolution of the unstable axisymmetric ring from Fig.
\ref{fig12} (left panel), and outcome of the evolution are
displayed in Fig. \ref{fig13}. It is observed that the system
tends to form a stationary necklace-like pattern composed of six
peaks towering above the remaining quasi-uniform background.

\begin{figure}[tbh]
\centerline{
\includegraphics[width=8cm,height=6cm,clip]{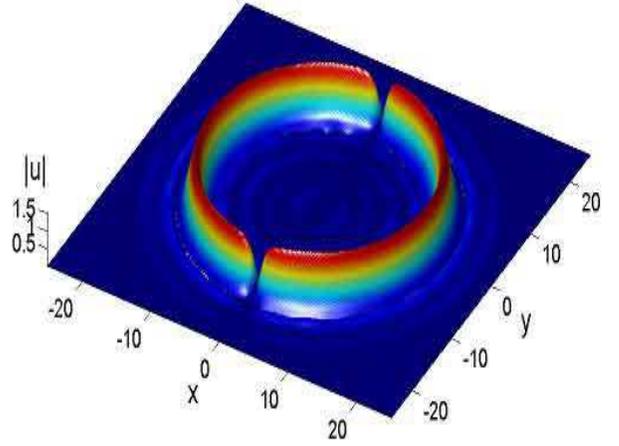}
} \caption{An example of a stable configuration formed by two dark
solitons created on top of a ring-shaped soliton in the repulsive
model with $\protect\varepsilon =3$, trapped at
$r_{0}=5\protect\pi $. This state has norm $N=330.6$ and chemical
potential $\protect\mu =0.80$.} \label{fig14}
\end{figure}

In the repulsive model, a stable ring (i.e., the radial gap
soliton) may additionally carry, on its crests, pairs of dark
solitons. An example of such a pattern is displayed in Fig.
\ref{fig14} (it resembles the dipole-mode ring soliton recently
found in the model with repulsion and Bessel lattice in Ref.
\cite{Ricardo}).

\begin{figure}[tbh]
\centerline{\includegraphics[width=8cm,height=6cm,clip]{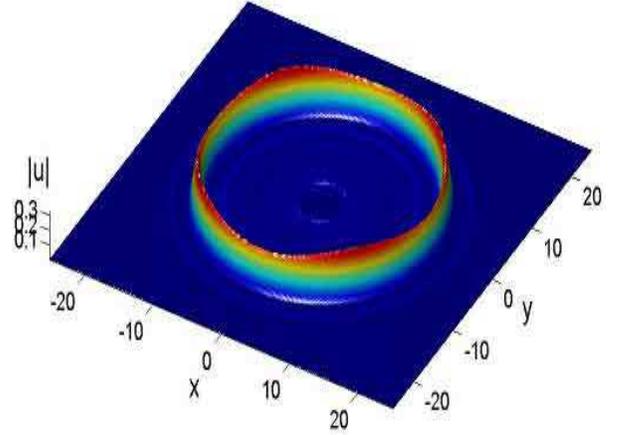}}
\caption{A stable ring-shaped soliton featuring weak azimuthal
modulation, in the model with attraction and strong radial lattice
($\protect\varepsilon =10$). The solution's norm and chemical
potential are $N=7.2$ and $\protect\mu =-5.7$.} \label{fig15}
\end{figure}

Stable static rings with a density modulation in the azimuthal
direction, corresponding to analytical solutions (\ref{cn}) for
the cnoidal waves, have been found too in the attractive model. A
typical example of such a pattern is displayed in Fig.
\ref{fig15}; in fact, it was generated as an outcome of the
evolution of an unstable axisymmetric ring with the same norm.

\begin{figure}[tbh]
\centerline{\includegraphics[width=8cm,height=6cm,clip]{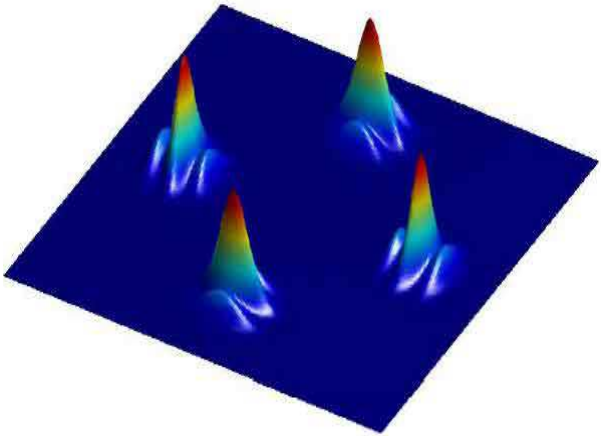}}
\caption{A stable set of four solitons in the attractive model,
trapped in a circular trough with $r_{0}=5\protect\pi $ and
$\protect\varepsilon =2$. The norm of each soliton is
$N=2\protect\pi $, with the phase difference $\protect\pi $
between adjacent ones. The lattice is embedded in an integration
domain of size $16\protect\pi \times 16\protect\pi $. }
\label{fig16}
\end{figure}

Increase of the norm leads to a transition from weakly modulated
stable patterns to deeply modulated ones. An example of such a
stable configuration, generated from four Gaussian pulses, with
the phase shifts $\pi $ between them, by integration of Eq.
(\ref{gpe2D}) in imaginary time, is displayed in Fig. \ref{fig16}
(note that the total norm of this pattern exceeds that
corresponding to Fig. \ref{fig15} by a factor of $\simeq 3.5$).

\section{Rotational dynamics and collisions between solitons}

\subsection{Stability of rotary solitons}

\begin{figure}[tbh]
\centerline{\includegraphics[width=8cm,height=8cm,clip]{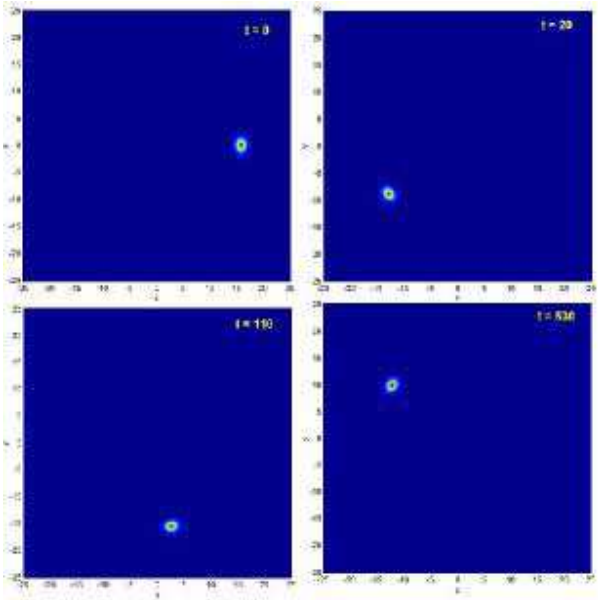}}
\caption{A stable soliton in the attractive model, with norm
$N=9.42$, trapped in the circular trough at $r_{0}=5\protect\pi $,
was set in motion in the clockwise direction with velocity $v=1$.
As illustrated by the sequence of snapshots, the circular motion
in the trough continues indefinitely long without any visible
loss.} \label{fig17}
\end{figure}

Strongly localized solitary waves of the attractive BEC,
self-trapped in a large-radius annular potential channel, like
those shown in Fig. \ref{fig16}, can freely move along in the
channel. If set in motion with a sufficiently small velocity, the
soliton can circulate in the channel indefinitely long, preserving
its integrity, as shown in Fig. \ref{fig17}.

\begin{figure}[tbh]
\centerline{\includegraphics[width=8cm,height=8cm,clip]{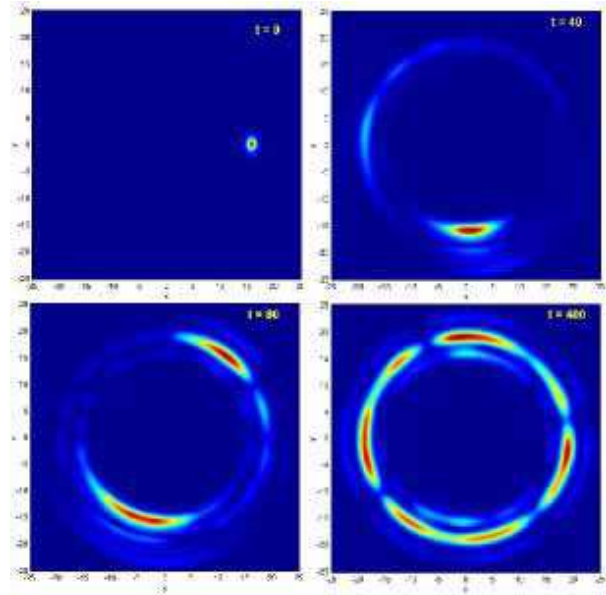}}
\caption{If the same soliton as in Fig. \protect\ref{fig17} (with
initial norm $N=9.42$) is set in motion with a greater velocity,
$v=3$, the centrifugal force gives rise to the underbarrier
leakage of atoms into the outer circular troughs. As illustrated
by the sequence of snapshots, the decaying soliton cannot preserve
its shape. The norm remaining within the given radial channel at
$t=400$ is $N_{\mathrm{final}}\simeq 0.5$, i.e., $\simeq 5\%$ of
the initial value.} \label{fig18}
\end{figure}

However, when the circulation speed is too large, the centrifugal
force acting on the soliton can cause the matter to tunnel into
adjacent radial troughs. The loss of the norm (number of atoms)
caused by the centrifugal tunneling eventually brings the
soliton's norm below the threshold value necessary for its
existence, which leads to disintegration of the soliton, see Fig.
\ref{fig18}.

If the lattice is weak, and the soliton is strongly self-trapped (with a
sufficiently large norm), the centrifugal force can also drive it away from
the center, across the lattice. In that case (not shown here), the soliton
also suffers the radiation loss, and eventually disintegrates.

\subsection{Collisions between solitons}

The present model offers unique possibilities for exploring
interactions and collisions between matter-wave solitons, as the
dynamics in circular channels is free of external perturbations,
which are always present in previously reported settings (in the
form of the weak longitudinal trap intended to keep the solitons
within a finite spatial domain). The same advantage is offered by
setups with the toroidal trap
\cite{torus-experiment,torus-theory}. However, sideline
(tangential) collisions, which are possible between solitons
moving in adjacent potential troughs of the radial lattice, have
no counterparts in the toroidal traps.

\begin{figure}[tbh]
\centerline{
\includegraphics[width=4cm,height=4cm,clip]{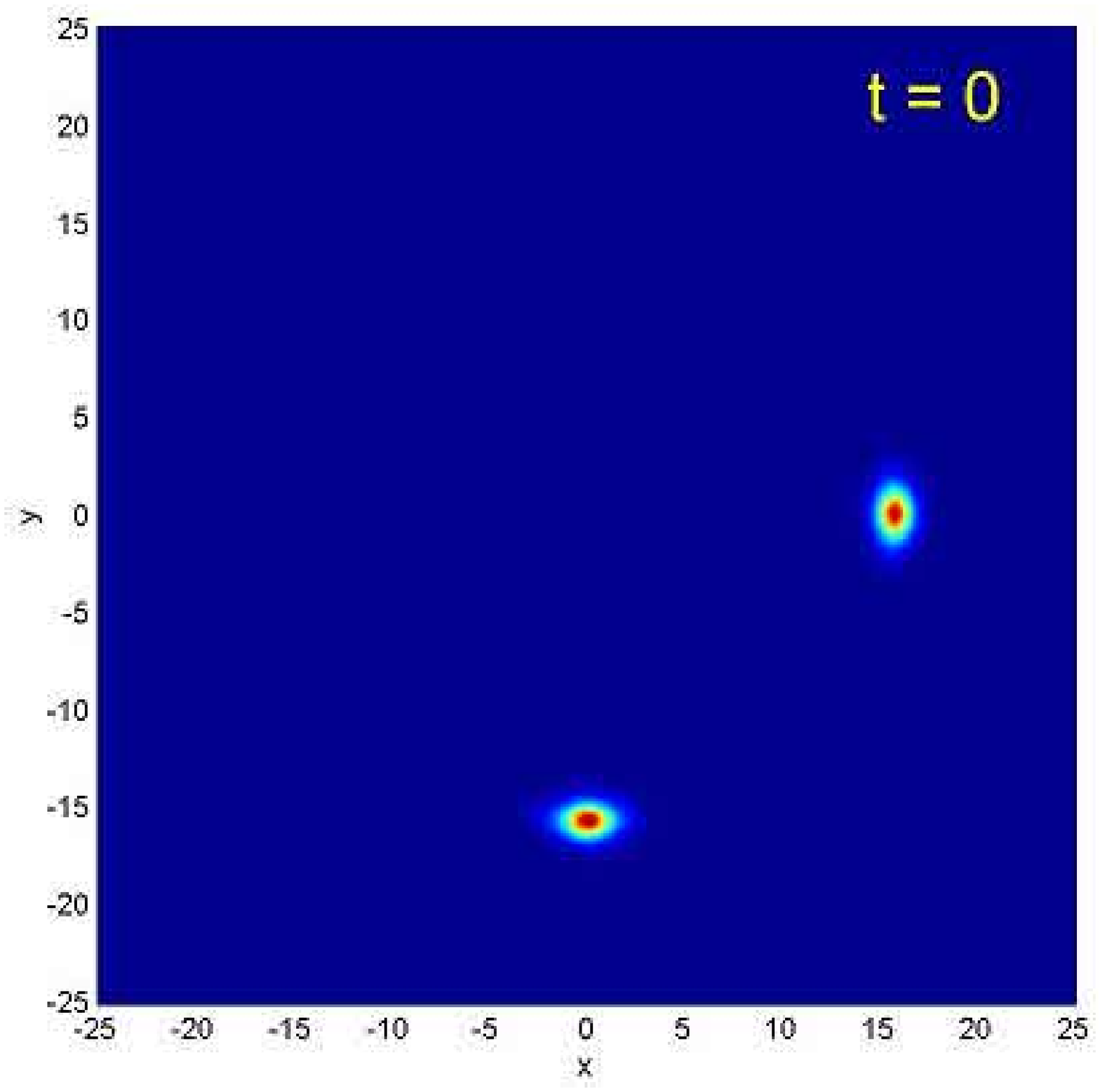} \quad
\includegraphics[width=4cm,height=4cm,clip]{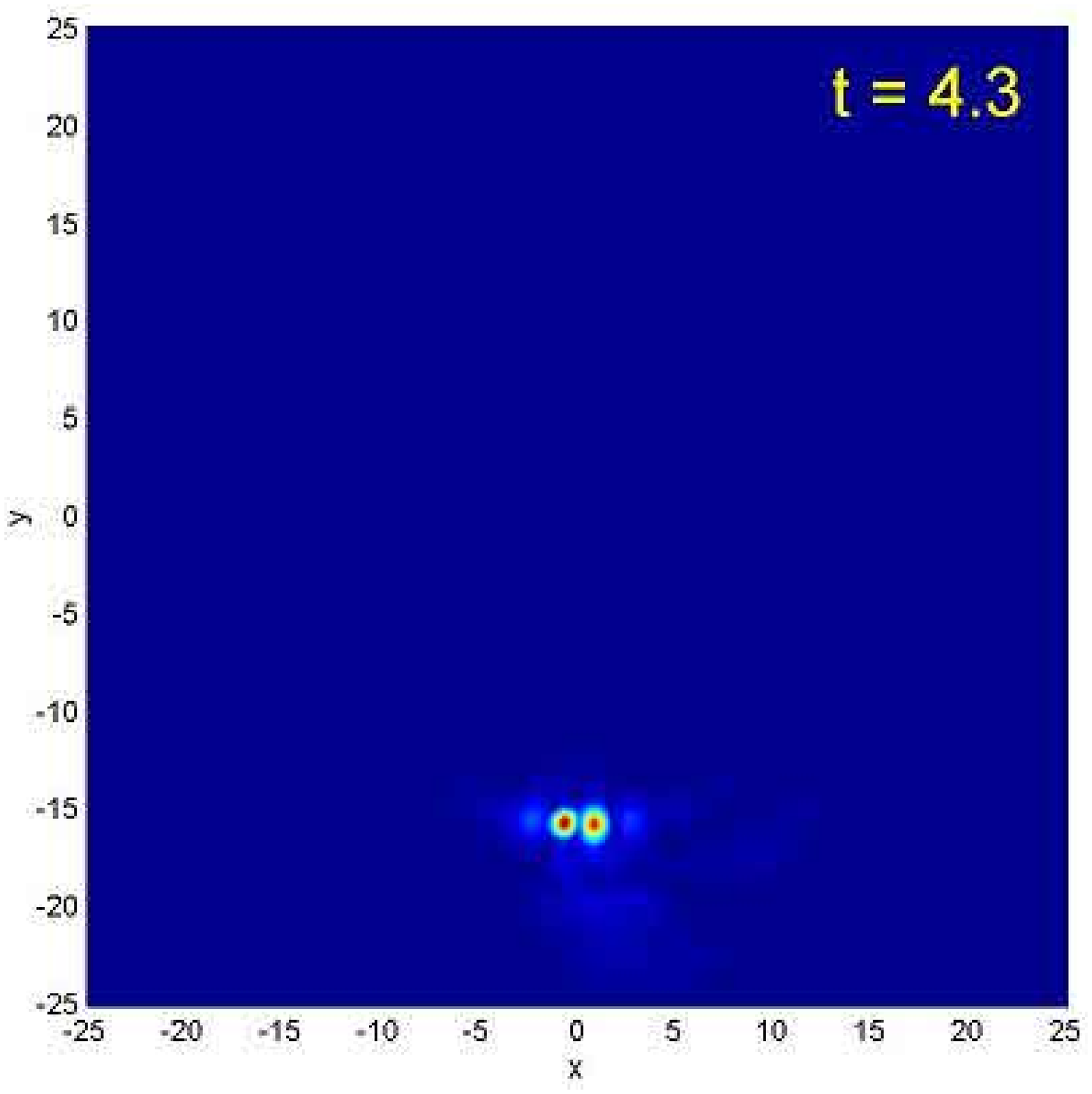}}
\centerline{
\includegraphics[width=4cm,height=4cm,clip]{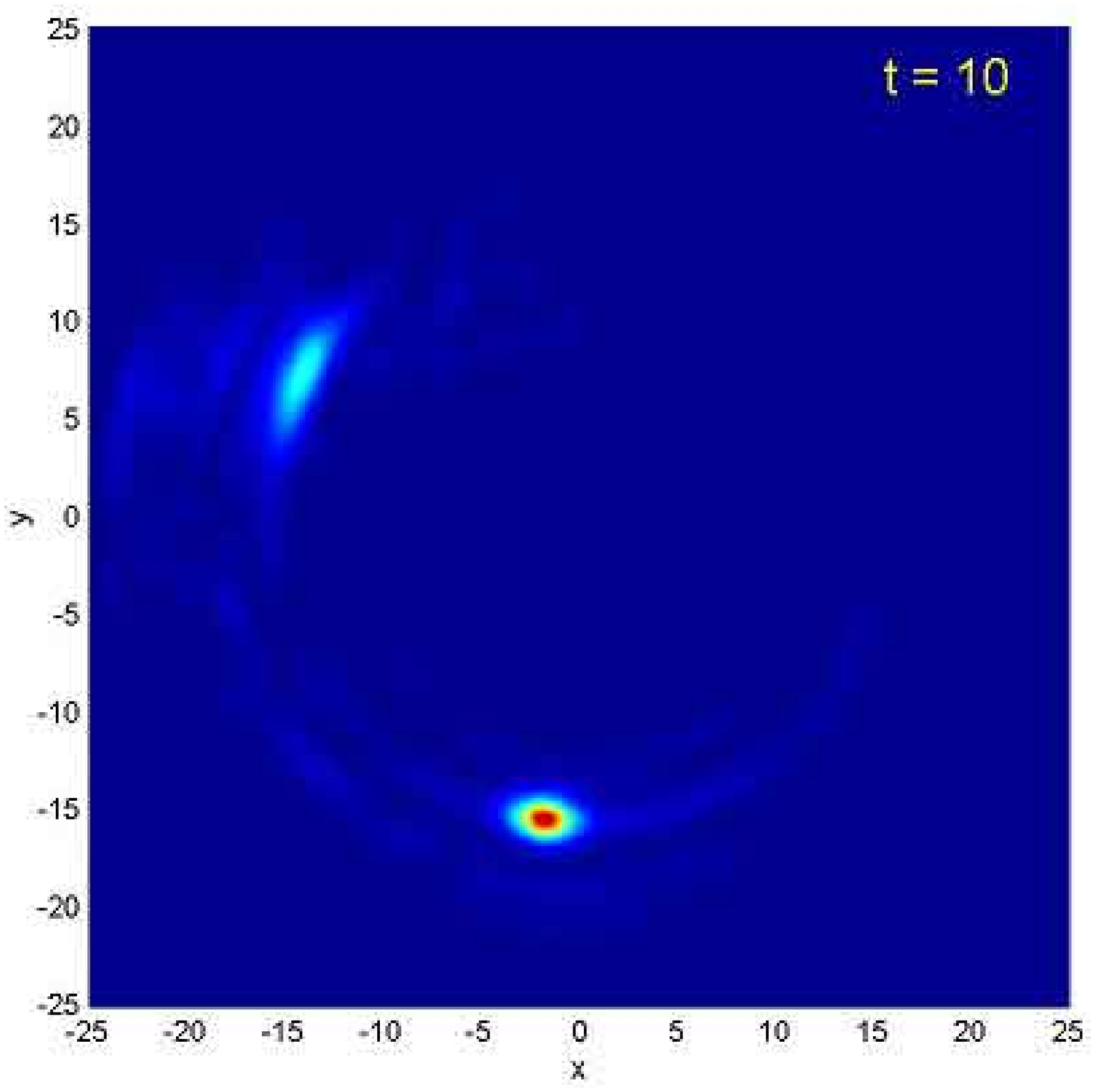} \quad
\includegraphics[width=4cm,height=4cm,clip]{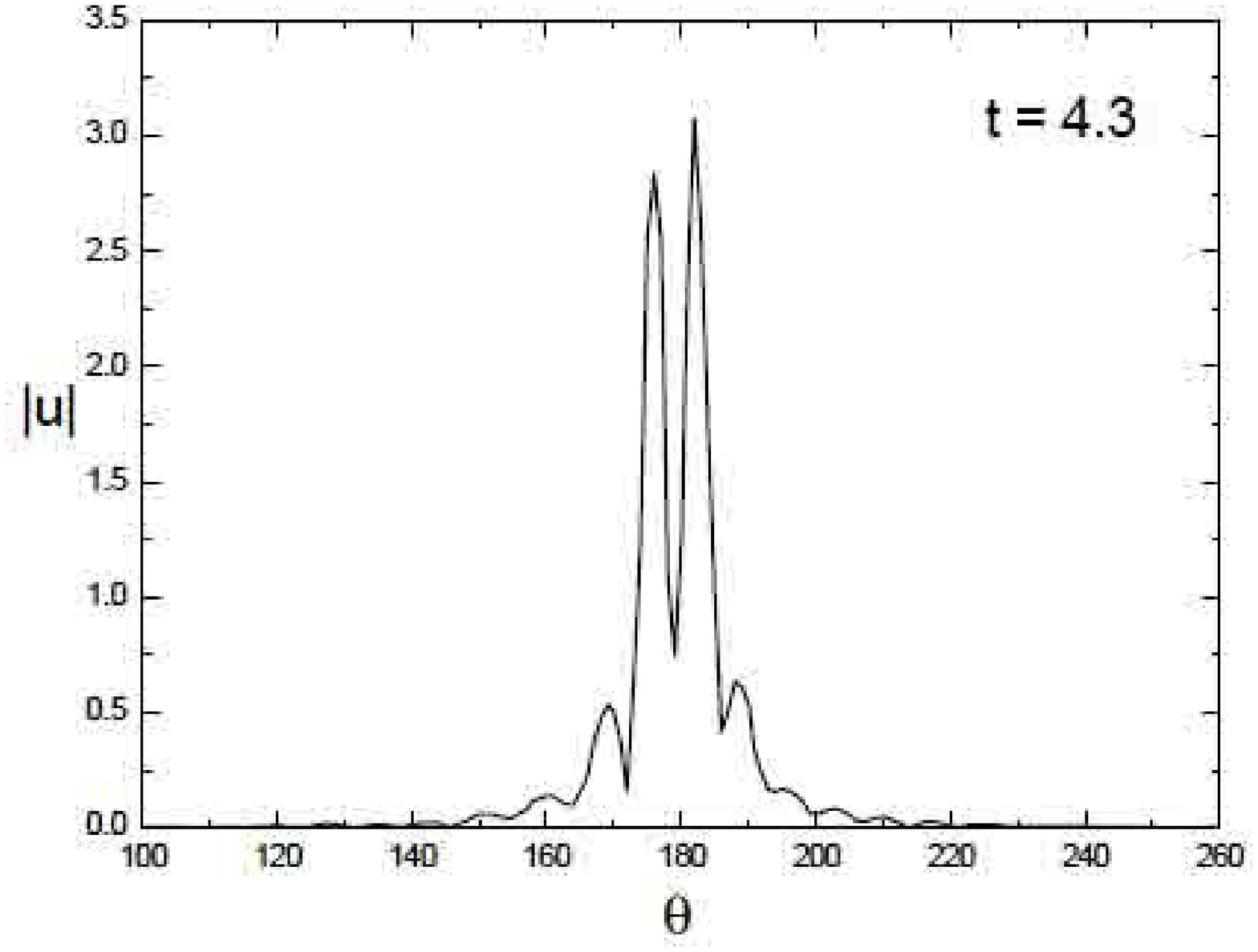}}
\caption{Collision between two solitons with individual norms
$N=2.5\protect\pi $ and zero phase difference, in the circular
trough at $r_{0}=5\protect\pi $. One of the solitons is set in
motion (at $x=5\protect\pi ,y=0$) at $t=0$ with speed $v=3$ in the
clockwise direction, while the second one (at $x=0,y=-5\protect\pi
$) stays quiescent. The norm remaining within the given circular
channel at $t=10$ is $N_{\mathrm{final}}=12.75$. The soliton which
was originally at rest loses less norm than the moving one. }
\label{fig19}
\end{figure}

\begin{figure}[tbh]
\centerline{
\includegraphics[width=4cm,height=4cm,clip]{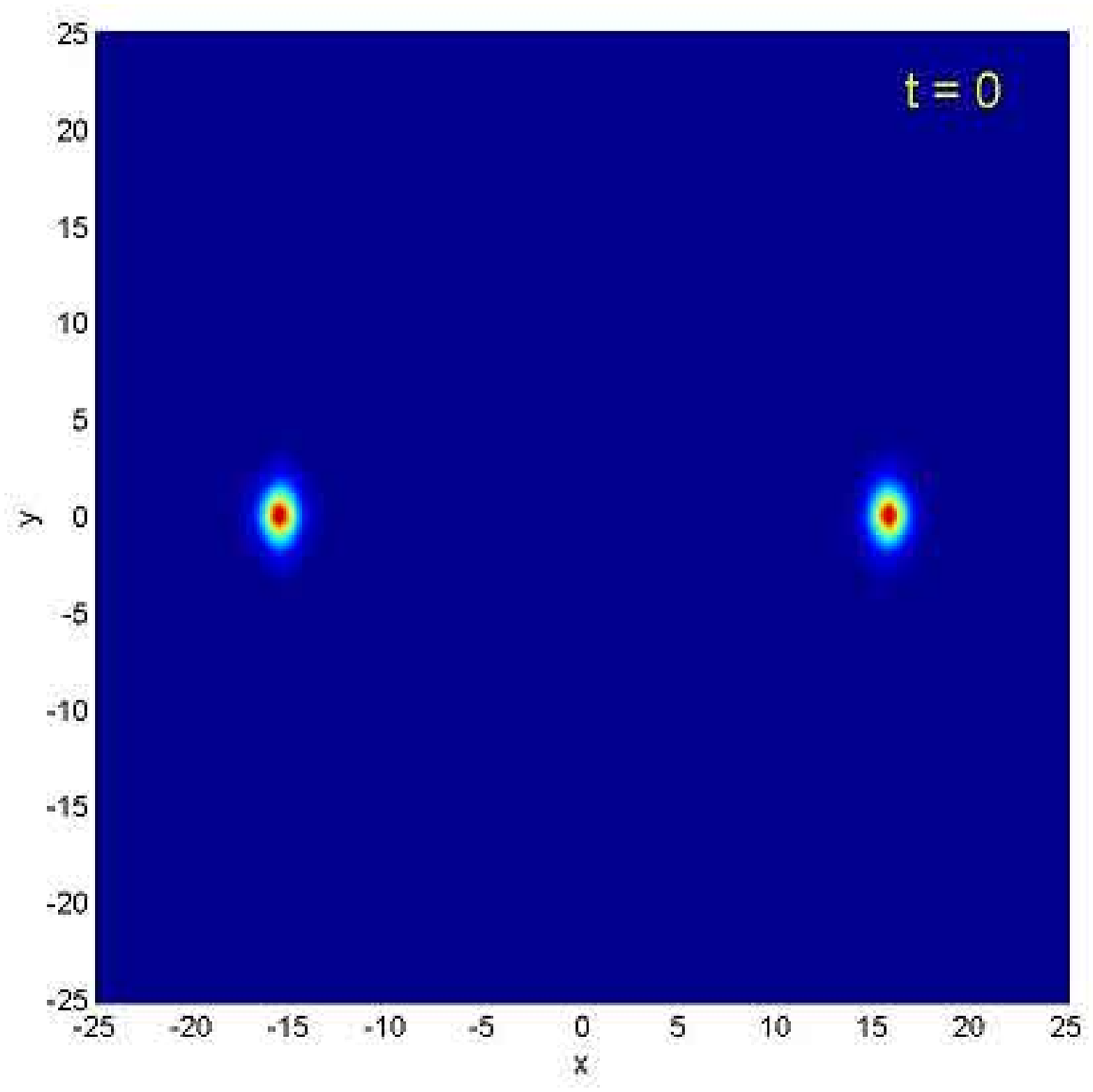} \quad
\includegraphics[width=4cm,height=4cm,clip]{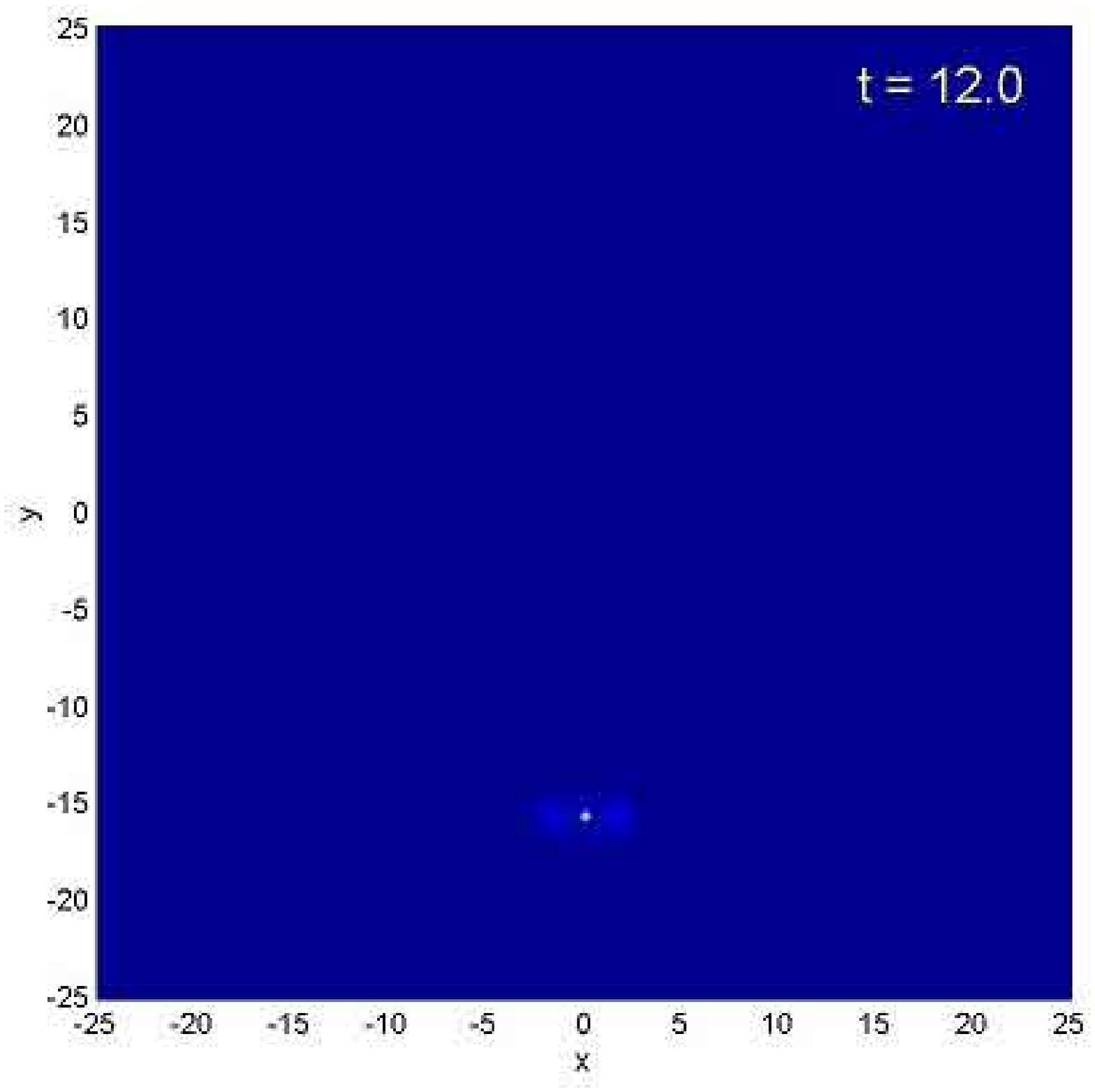}}
\centerline{\includegraphics[width=8cm,height=6cm,clip]{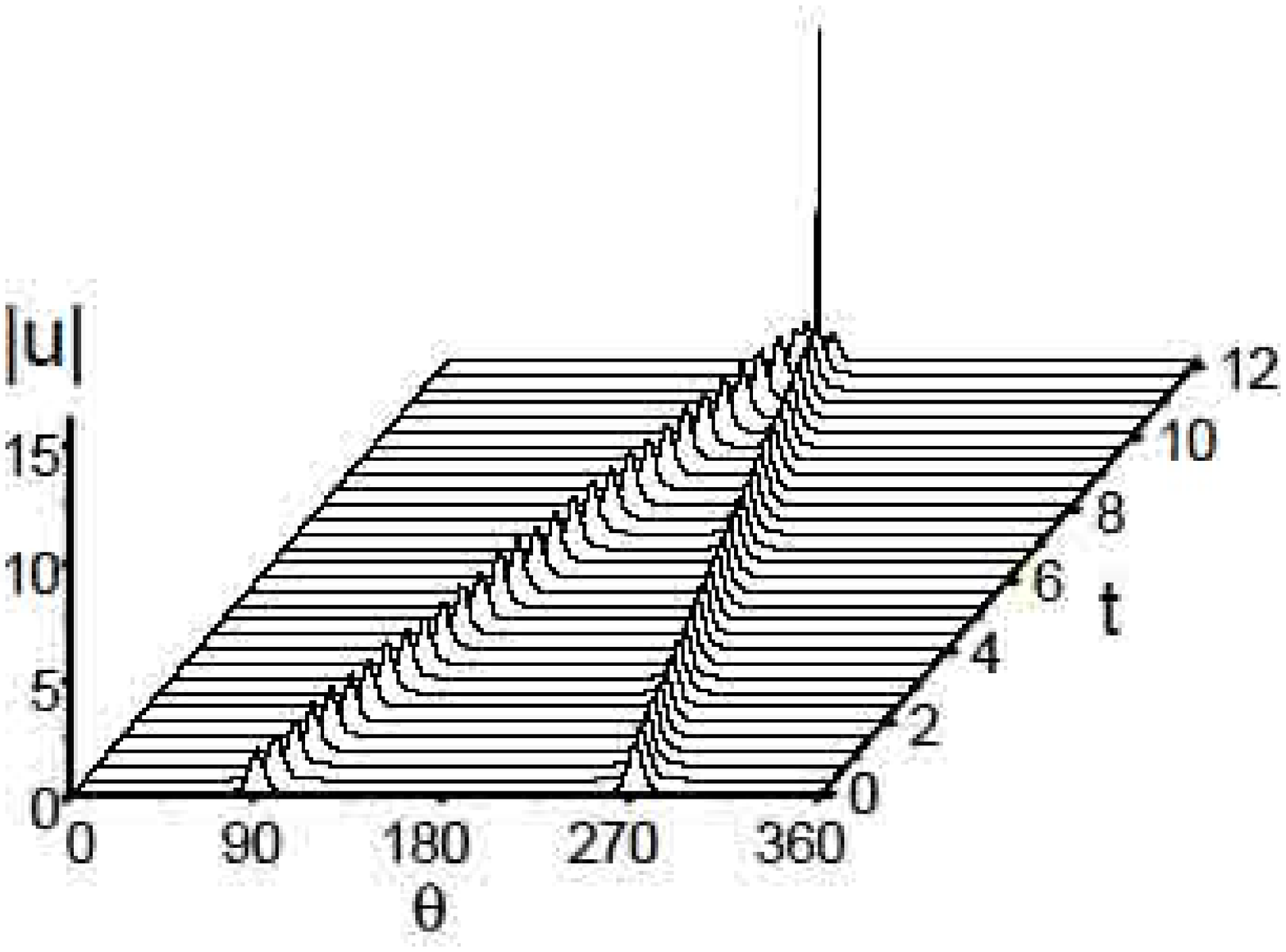}}
\caption{The head-on collision of two in-phase solitons in the
attractive model, when both of them were set in motion with
relatively small velocities, $v=\pm 1$, in one circular channel
($r_{0}=5\protect\pi $), ends up with collapse. Upper panels:
density plots at $t=0$ and the moment of collapse ($t=12$). Lower
panel: the evolution of the density profiles in the cross section
along the circumference of the potential trough. As in Fig.
\protect\ref{fig19}, the initial norm of each soliton is
$N=2.5\protect\pi $. } \label{fig20}
\end{figure}

In head-on collisions of solitons in the attractive model (in the
same circular channel), two competing processes play major roles.
The first is the nonlinear self-focusing, which may give rise to
intrinsic collapse of a large-norm ``lump" temporarily formed by
the colliding solitons, and the other is the interference due to
coherence of the matter-wave solitons \cite{bms2004,Estoril}. The
interference pattern, featuring alternating regions of high and
low density, can suppress the collapse. The predominance of either
mechanism depends on the time of interaction. If it is small (in
the case of large collision speed), solitons can pass trough each
other without collapse (although with conspicuous loss, see below,
hence the collision cannot be termed quasi-elastic), even if their
total norm exceeds the collapse threshold. On the contrary, slow
collisions lead to the onset of collapse. These two possible
scenarios of the head-on collisions between in-phase (mutually
attracting) solitons are shown in Figs. \ref{fig19} and
\ref{fig20}.

Figure \ref{fig19} presents an example of a relatively fast
collision. At the moment of the full overlap between the solitons,
$t=4.3$, the interference fringes are evident, and are also seen
in the cross section along the circumference of the channel (lower
right panel). Although the total norm of the colliding solitons is
overcritical ($N_{\mathrm{tot}}=14.9$), and the collision time
($\Delta t\simeq 0.5$) is larger than the collapse time (which is
estimated as $t_{\mathrm{collapse}}\sim 1/N_{\mathrm{tot}}$) , the
solitons separate without collapse. Nevertheless, the collision is
essentially inelastic, the loss being additionally enhanced by the
matter leakage under the action of the centrifugal force. On the
other hand, Fig. \ref{fig20} shows that the solitons colliding at
a sufficiently small velocity merge together and indeed blow up
due to the collapse.

\ \

\begin{figure}[tbh]
\centerline{\includegraphics[width=6cm,height=6cm,clip]{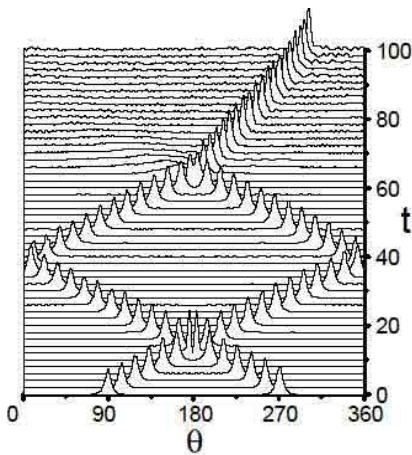}}
\caption{Collision of two solitons with the same parameters as in
Fig. \protect\ref{fig20}, but the phase difference of $\protect\pi
$ between them. Solitons bounce at $\protect\theta
=180^{\mathrm{o}}$ and $\protect\theta =0^{\mathrm{o}}$. Finally,
the solitons merge into a single one, without collapse.}
\label{fig21}
\end{figure}

Two $\pi $ out-of-phase solitons colliding in the same potential
trough bounce from each other due to repulsion between them. Then,
repeated bouncing collisions are observed, which is similar to the
dynamics in the Bessel lattice \cite{kartashov}, as well as in a
quasi-1D OL, equipped with a weak longitudinal trap
\cite{bms2004}. However, in contrast to those works, we have
observed that many collisions in the circular trough gradually
wash out the phase coherence between the solitons. As a result,
the solitons cease to repel each other, and a clear interference
pattern disappears. Eventually, the originally repelling solitons
merge together, shedding the excessive norm away, as shown in Fig.
\ref{fig21}. The intrinsic collapse \emph{does not} occur in this
case, as the loss makes the norm of the finally established single
soliton smaller than the critical value at the collapse threshold.

\begin{figure}[tbh]
\centerline{
\includegraphics[width=3cm,height=3cm,clip]{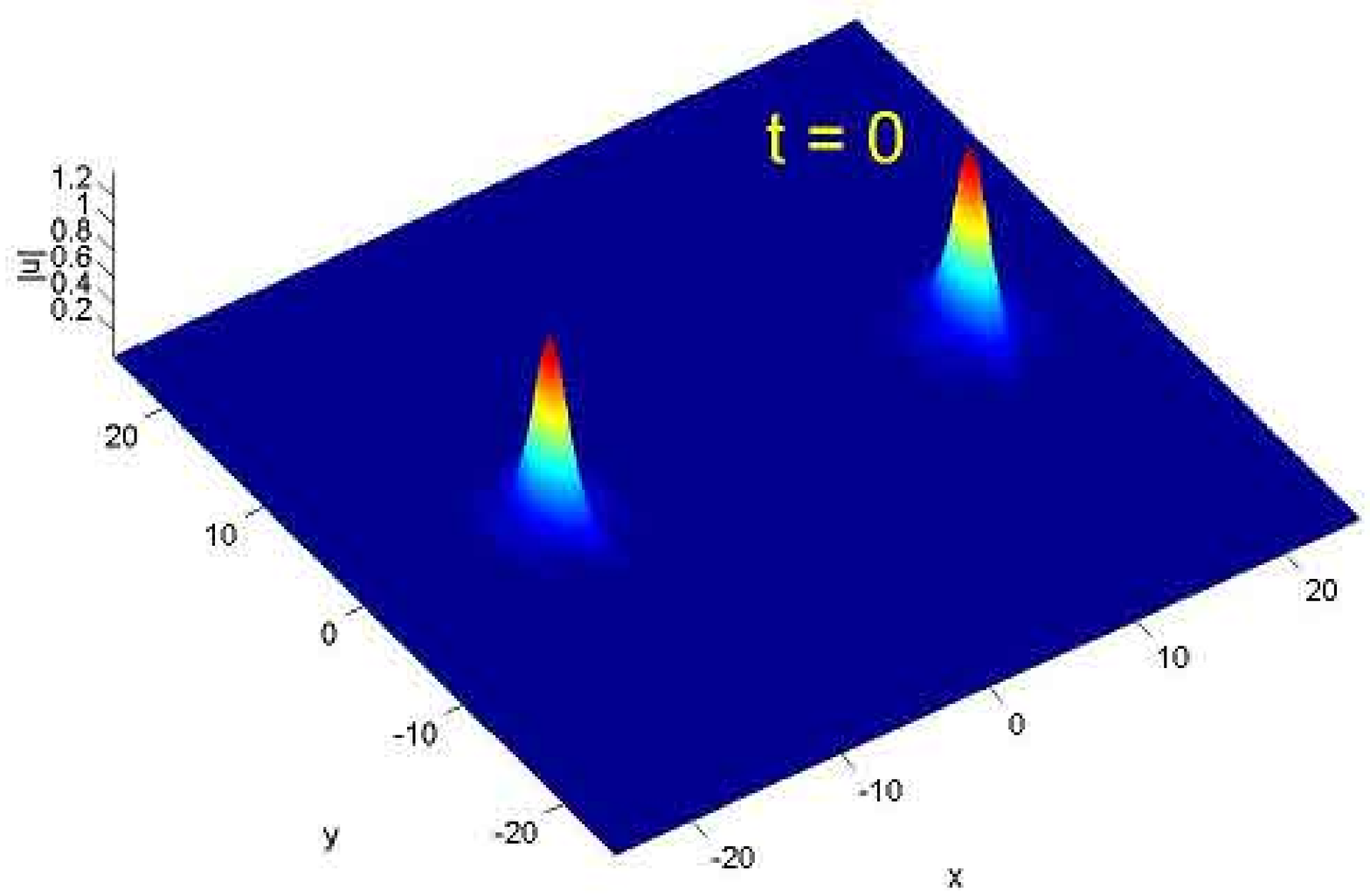}
\includegraphics[width=3cm,height=3cm,clip]{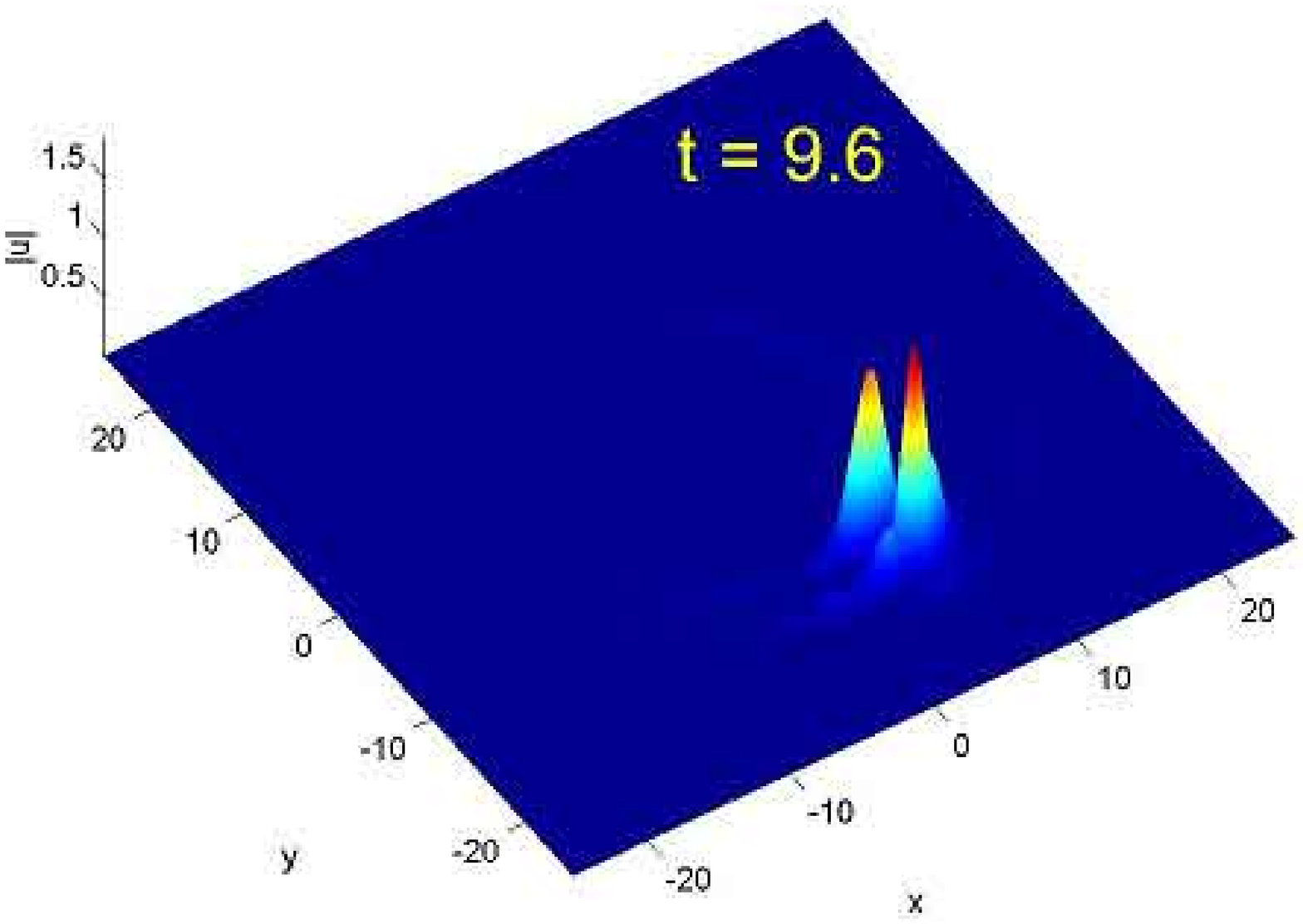}
\includegraphics[width=3cm,height=3cm,clip]{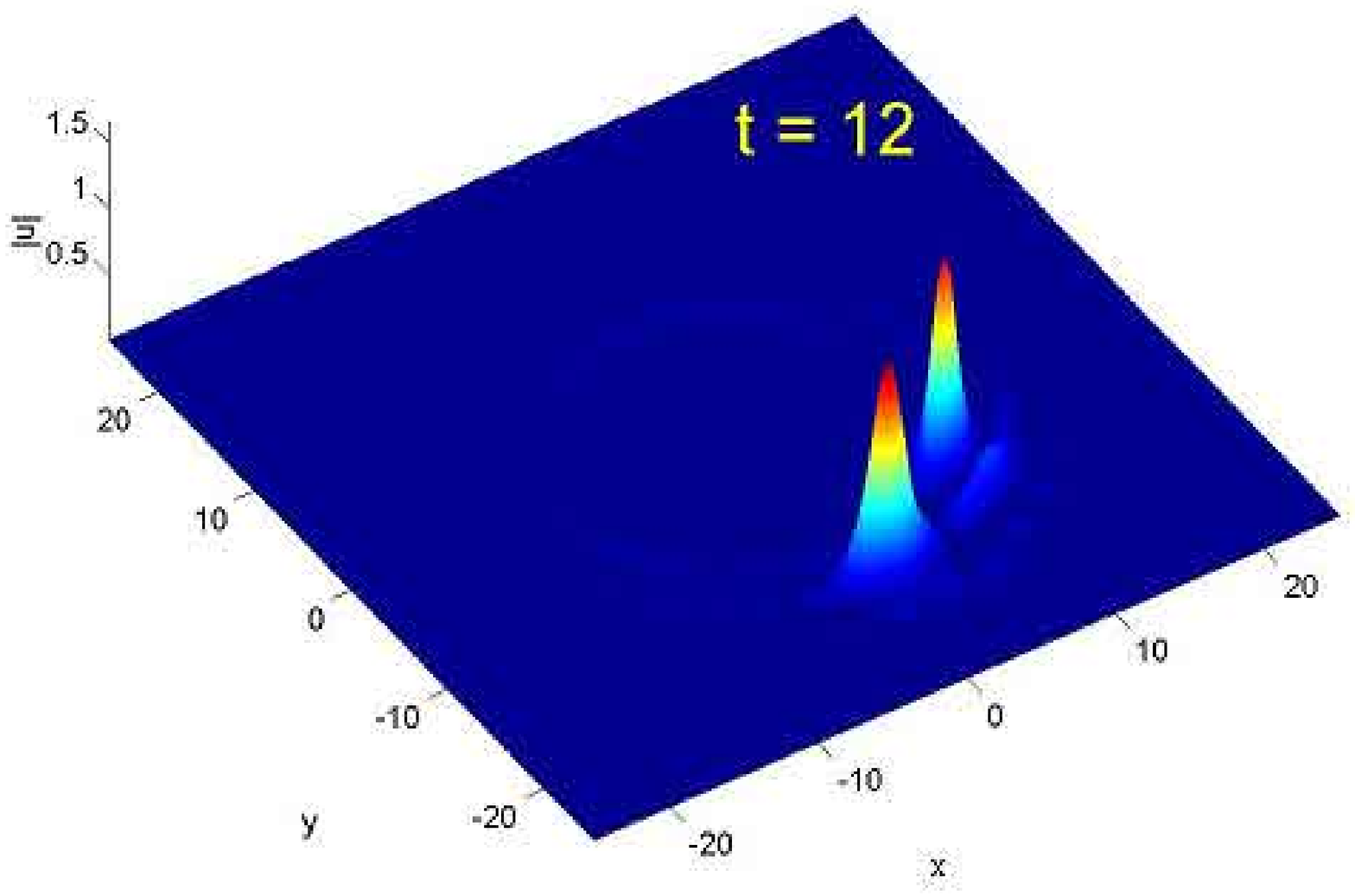}}
\caption{A sequence of snapshots for the collision between
solitons with individual norms $N=2\protect\pi $, moving at
velocities $v=\pm 1$ in adjacent troughs (at $r_{0}=5\protect\pi $
and $r_{0}=4\protect\pi $) of the radial lattice with
$\protect\varepsilon =2$. The solitons survive the collision, and
remain localized, although the collision is inelastic, giving rise
to radiation loss. } \label{fig22}
\end{figure}

Collisions of matter-wave solitons in the attractive BEC is
accompanied by strong exchange of matter between them. The present
setting makes it possible to investigate related phenomena for
solitons moving in adjacent channels of the radial potential, and
thus colliding tangentially. The result is that two in-phase
solitons are unstable against the flow of matter in this setting,
and tend to merge (a similar trend was observed in the model with
a quasi-1D lattice in Ref. \cite{bms2004}). Nevertheless, the
solitons may avoid collapse and separate, if the collision time is
small due to a large relative velocity, as shown in Fig.
\ref{fig22} .

\begin{figure}[tbh]
\centerline{
\includegraphics[width=3cm,height=3cm,clip]{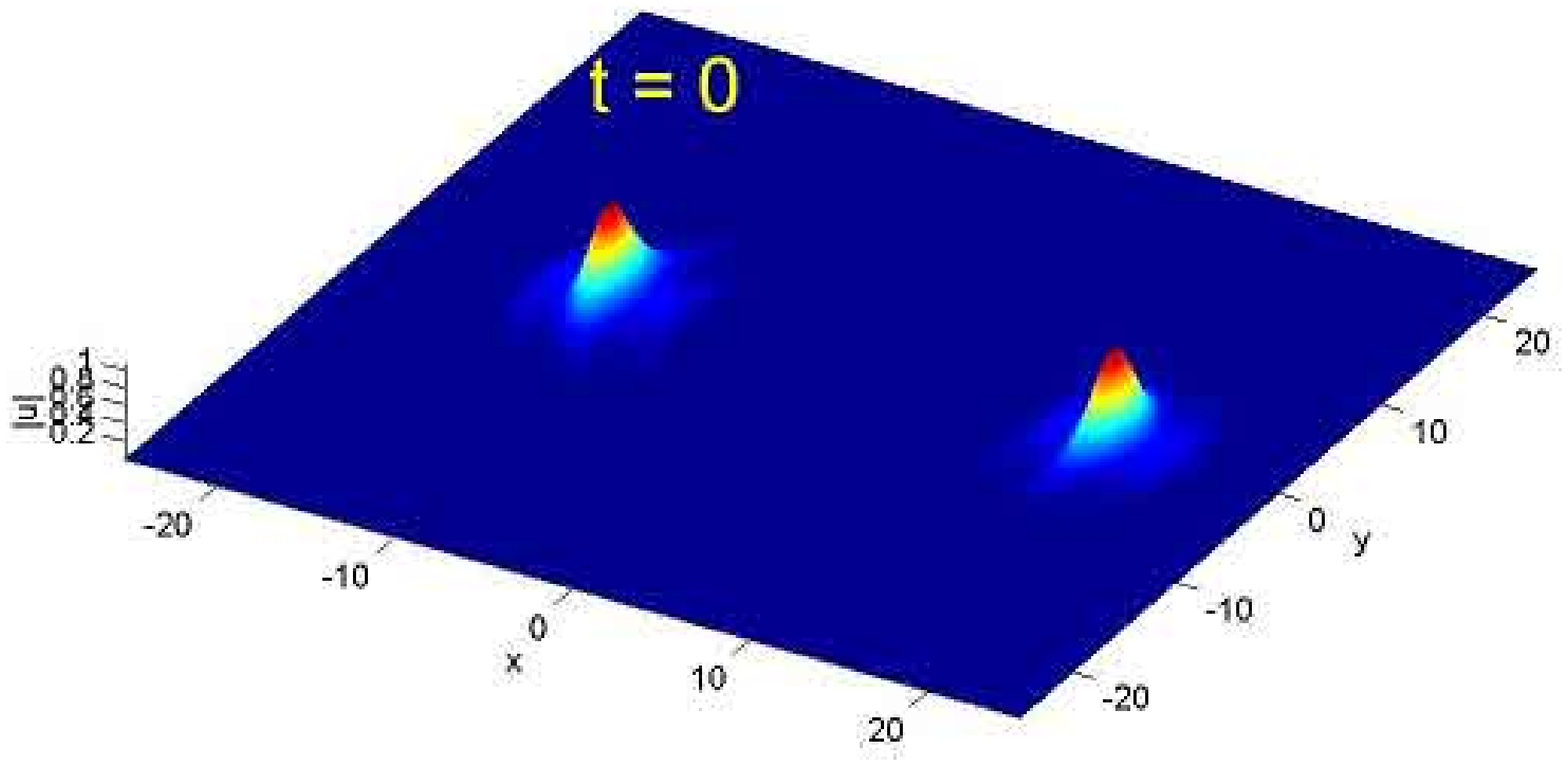}
\includegraphics[width=3cm,height=3cm,clip]{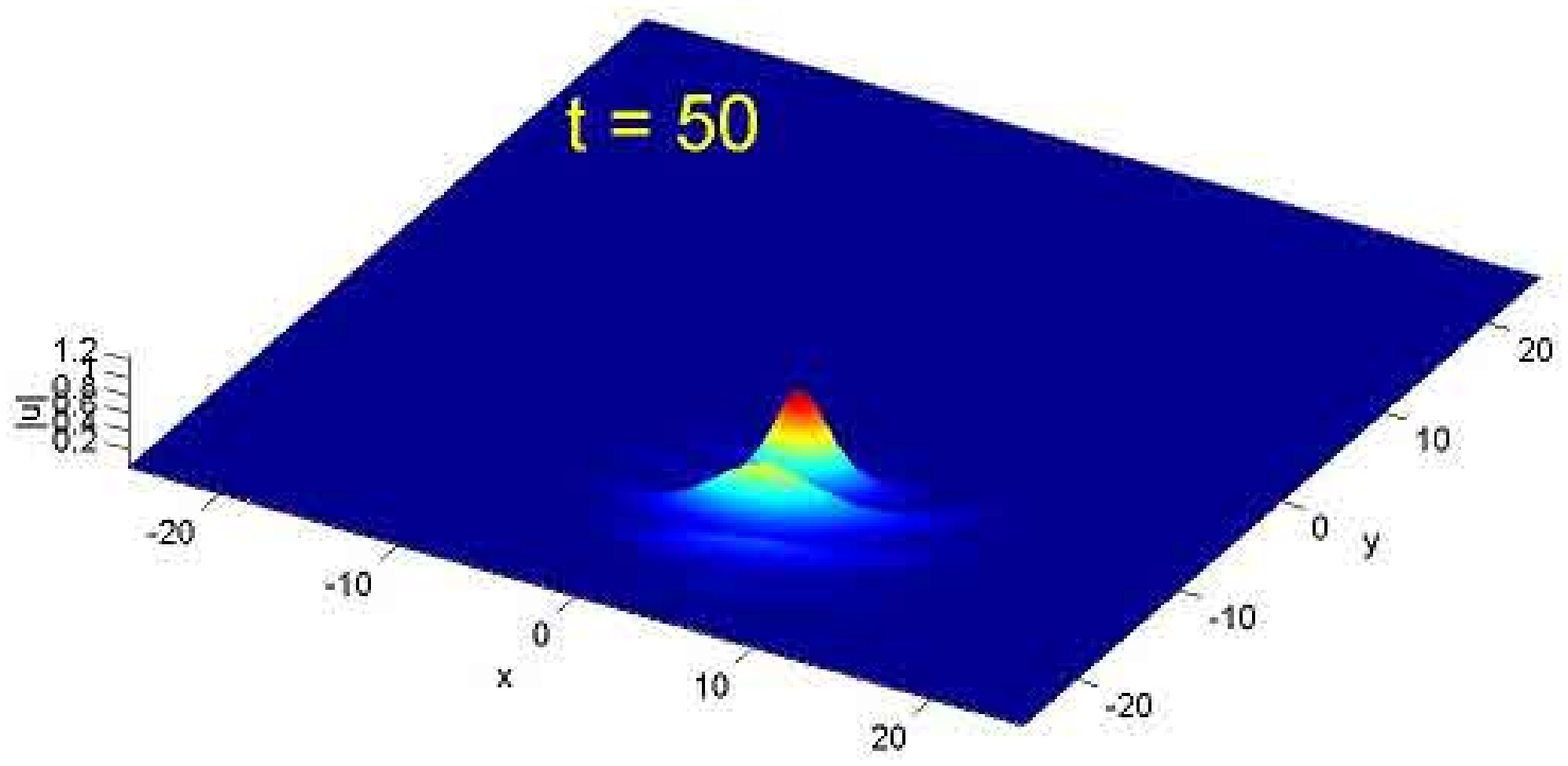}
\includegraphics[width=3cm,height=3cm,clip]{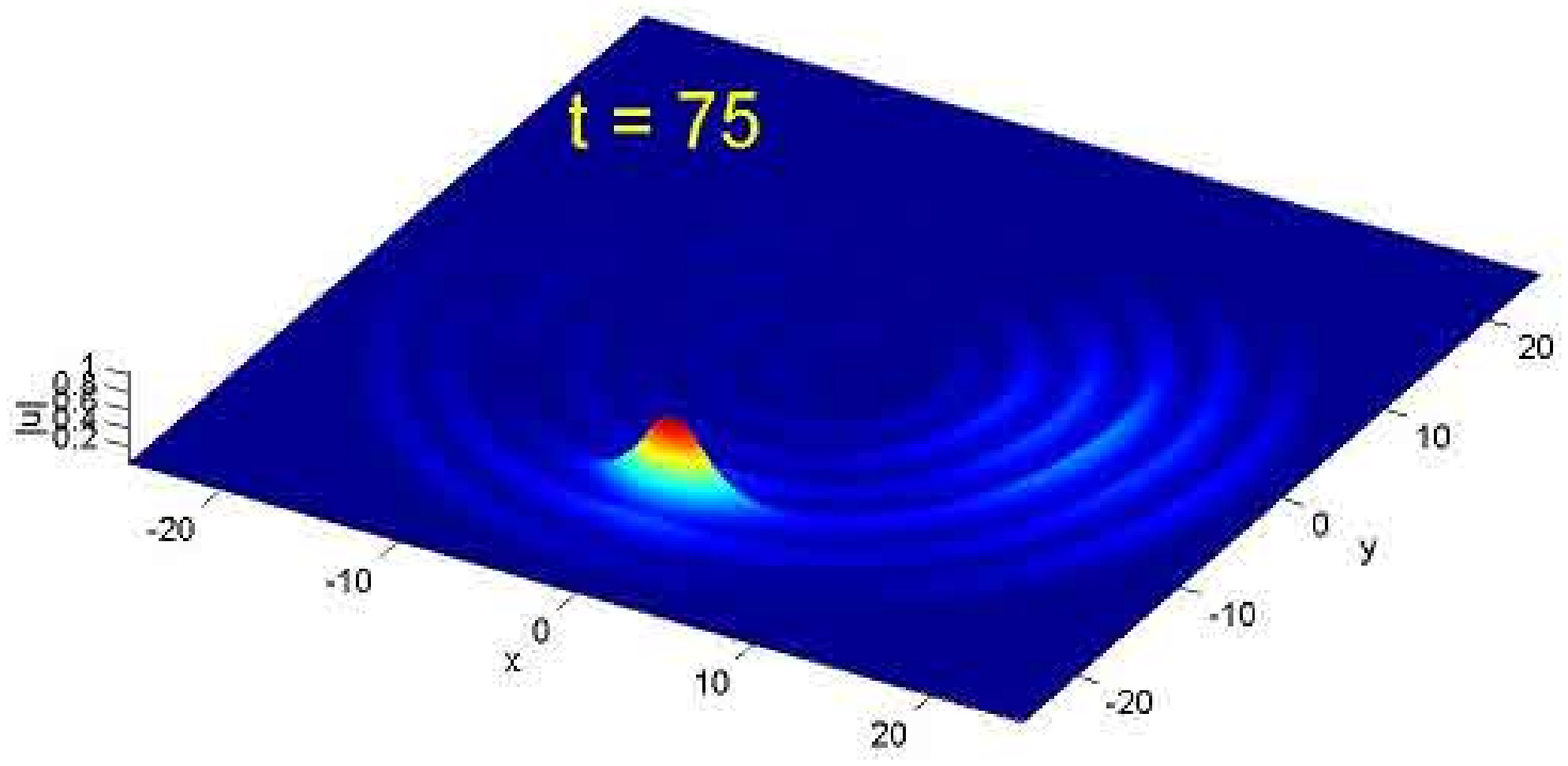}}
\caption{The same as in Fig. \protect\ref{fig15}, but for solitons
with smaller norms, $N=1.7\protect\pi $, launched to move at
velocities $v=\pm 0.2 $.} \label{fig23}
\end{figure}

On the other hand, if the solitons are loosely bound (i.e. their
norms are small and/or the radial lattice is weak), the slow
lateral collision gives rise to merger of the solitons into one,
excessive norm being shed off with linear waves, as demonstrated
in Fig. \ref{fig23}.

\section{Conclusions}

In this work, we have proposed a new setting to explore
two-dimensional (2D) localized states in self-attractive and
repulsive BECs, in the form of a periodic radial optical lattice.
A crucial difference from previously studied 2D models with Bessel
lattices is the fact that no localized state may exist in the
linear limit of the present model, hence all solitary states in it
are ``true solitons", impossible without the nonlinearity. Besides
the BEC, the model may also apply to spatial solitons (beams) in
photonic crystal fibers with a circular intrinsic structure.

The existence of such objects was demonstrated by means of different
variants of the variational approximation, and in direct numerical
simulations. We have investigated the localized states trapped in the
central potential well and in remote circular potential troughs. In both
cases, a new species, namely, stable \textit{radial gap solitons}, have been
identified in the model with self-repulsion (in addition, the radial gap
soliton trapped in a circular trough can carry stable pairs of dark solitons
on its crest). In remote troughs, we have investigated ring-shaped patterns
delocalized in the azimuthal direction, as well as strongly localized
azimuthal solitons (including moving ones). In the attractive model,
solutions of both types are described by an effective 1D equation
supplemented by the periodic boundary conditions (it is a nonpolynomial NLS
equation with the coordinate running along the circumference). In
particular, a finite threshold of the azimuthal modulational instability
(MI)\ of axisymmetric ring-shaped states was predicted, and exact
cnoidal-wave solutions, generated by the MI-induced bifurcation were found
in the framework of the 1D equation. Azimuthal solitons were also found as
solutions to the same equation. The existence of stable azimuthally uniform
and weakly modulated ring-shaped states was corroborated by direct
simulations.

Dynamics of completely localized solitons circulating in the annular
potential troughs was investigated in the attractive model by means of
direct simulations. In particular, the solitons with sufficiently small
velocities remain stable indefinitely long, while high velocities give rise
to leakage of matter into the adjacent (more remote) trough under the action
of the centrifugal force, which eventually destroys the soliton. Collisions
between solitary waves running in the same or adjacent circular channels
were investigated too. Head-on collisions of the in-phase solitons in one
trough lead to the collapse; $\pi $-out of phase solitons bounce from each
other many times, but gradually loose their mutual coherence, and eventually
merge into a single soliton without collapsing, shedding off excess norm.
In-phase solitons colliding in adjacent channels may also merge into a
single soliton.

\section*{Acknowledgements}

B.B.B. is partially supported by the grant No. 20-06 from the Fund
for Fundamental Research of the Uzbek Academy of Sciences. The
work of B.A.M. was supported, in a part, by the
Center-of-Excellence grant No. 8006/03 from the Israel Science
Foundation. MS acknowledges financial support from the MIUR,
through the inter-university project:{\it Dynamical properties of
Bose-Einstein condensates in optical lattices}, PRIN-2005.

\end{document}